\documentclass[twocolumn,showpacs,superscriptaddress,prb,amsmath,amssymb,floatfix,eqsecnum]{revtex4-1}
\usepackage{amsmath,amssymb}
\usepackage{graphicx}
\usepackage[latin1]{inputenc}
\usepackage{bm}
\usepackage{psfrag}
\usepackage{color}

\newcommand{\bd}{\bm}

\newcommand{\nspace}{
\vspace{-.3cm}
}

\begin{document}

\title{Thermalization of magnons in yttrium-iron garnet: \\
nonequilibrium functional renormalization group approach}

\author{Johannes Hick}  
\affiliation{Institut f\"{u}r Theoretische Physik, Universit\"{a}t
 Frankfurt,  Max-von-Laue Strasse 1, 60438 Frankfurt, Germany}

\author{Thomas Kloss}  
\affiliation{
Laboratoire de Physique et Mod\'elisation des Milieux Condens\'es, CNRS and Universit\'e Joseph Fourier, 25 Avenue des Martyrs, 38042 Grenoble, France}

\author{
Peter Kopietz} 
\affiliation{Institut f\"{u}r Theoretische Physik, Universit\"{a}t
  Frankfurt,  Max-von-Laue Strasse 1, 60438 Frankfurt, Germany}

\date{August 09, 2012}

 \begin{abstract}

Using a nonequilibrium functional renormalization group (FRG)
approach we calculate the time evolution  of 
the momentum distribution of a magnon gas
in contact with a thermal phonon bath.
As a cutoff for the FRG procedure
we use a hybridization parameter $\Lambda$ giving rise to
an artificial damping of the phonons.
Within our truncation of the FRG flow equations
the time evolution of the magnon distribution is obtained from a
rate equation involving  cutoff-dependent nonequilibrium
self-energies, which in turn satisfy FRG flow
equations depending on cutoff-dependent transition rates.
Our approach goes beyond the Born collision approximation
and takes the feedback of the magnons on the phonons into account.
We use our method to calculate the thermalization of a quasi two-dimensional
magnon gas 
in the magnetic insulator yttrium-iron garnet after a
highly excited  initial state has been generated by an external microwave field.
We obtain good agreement with recent experiments.

\end{abstract}


\maketitle

\section{Introduction}

Calculating the nonequilibrium time evolution of interacting many-body systems
is a challenging problem which usually requires advanced
many body methods and serious numerical calculations.
This problem is of interest in
many different fields \cite{Morawetz04}, including ultracold atoms \cite{Polkovnikov11},
collision experiments of heavy nuclei \cite{Cassing09},  
exciton-polariton systems in semiconductor
microcavities \cite{Deng10}, dissipative quantum systems \cite{Schoeller09}
and  pumped magnon gases in magnetic 
insulators \cite{Demokritov06,Dzyapko11}.
The nonequilibrium dynamics of weakly interacting non-relativistic bosons
has been successfully modeled
by  combining the Gross-Pitaevskii equation for the condensate with the Boltzmann equation
for the quasi-particle excitations\cite{Griffin09,Kamenev11}.
However,  the collision integral in the Boltzmann equation 
is usually calculated only perturbatively to
second order in the bare interaction.
This approximation breaks down for strong interactions or for high densities, 
so that it is important to develop non-perturbative methods.
For $N$-component relativistic scalar field theories 
Berges and co-authors\cite{Berges09}  have recently shown by means of a
two-particle irreducible resummed large-$N$ expansion 
that  vertex corrections (which are neglected in the Boltzmann equation)
qualitatively change the scaling behavior of the momentum distribution
for small momenta. 

An alternative non-perturbative method to deal with 
strongly correlated systems out of equilibrium is based
on the functional renormalization group (FRG) \cite{Bagnuls01,Berges02,Kopietz10,Rosten12,Metzner12}.
Recently, several authors have used nonequilibrium
FRG methods to calculate  the time evolution of
various types of many body systems \cite{Gasenzer08,Kloss11,Kennes12}. 
It turns out, however, that the proper
implementation of  nonequilibrium FRG schemes 
depends on the specific problem of interest; in fact,  many technical problems 
(such as the choice of the cutoff parameter, or the construction
of sensible approximation strategies which respect causality and the
conservation laws)
have to be solved in order to turn the nonequilibrium FRG into a useful and competitive  
calculational tool.

In this work we shall develop a particular implementation of the
nonequilibrium FRG which is suited to
study time-dependent nonequilibrium phenomena in Bose systems
in contact with a thermal phonon bath. 
To be specific, we identify the bosons with the magnons (i.e., quantized spin-waves)
in a ferromagnetic insulator; however the methods developed in this work
are also useful to calculate the 
nonequilibrium time evolution of any non-relativistic Bose gas.
Moreover, by means of a  straightforward modification of our method 
it should also be possible to study Fermi gases out of equilibrium.
The Hamiltonian of our system is
 \begin{eqnarray}
 {\cal{H}} & = & \sum_{\bd{k}} \epsilon_{\bd{k}} a^{\dagger}_{\bd{k}} a_{\bd{k}} +
 \sum_{\bd{q}} \omega_{\bd{q}} b^{\dagger}_{\bd{q}} b_{\bd{q}} 
\nonumber
 \\
 & + & 
 \frac{1}{ \sqrt{V}} \sum_{\bd{q}} \gamma_{\bd{q}} \rho_{- \bd{q}} 
 ( b_{\bd{q}} + b^{\dagger}_{-\bd{q}} ),
 \label{eq:hdef}
 \end{eqnarray}
 where $a_{\bd{k}}^{\dagger}$ creates a magnon with momentum $\bd{k}$ and
energy $\epsilon_{\bd{k}}$, while 
$b^{\dagger}_{\bd{q}}$ creates an acoustic  phonon with momentum $\bd{q}$ and
energy $\omega_{\bd{q}} = c |\bd{q} |$. 
The volume of the system is denoted by $V$, 
the operator
$\rho_{\bd{q}} = \sum_{\bd{k}} a^{\dagger}_{\bd{k}} a_{\bd{k} + \bd{q}}$ 
represents Fourier components of the magnon density, and the magnon-phonon 
coupling $\gamma_{\bd{q}}$ is assumed to be proportional to
$\sqrt{\omega_{ \bd{q} }} $, which is a simple
consequence  of the fact that density fluctuations associated with the creation of
longitudinal phonons are proportional to the divergence of the 
corresponding displacement field\cite{Fetter71}.

 A model Hamiltonian of the form (\ref{eq:hdef}) has been used previously by
 B\'{a}nyai {\it{et al.}}\cite{Banyai00} to describe the  
kinetics of Bose-Einstein condensation of excitons.
In Sec.~\ref{sec:rate} we will review
the rate equation approach adopted by these authors. 
Here we use the model Hamiltonian (\ref{eq:hdef}) to study 
the nonequilibrium dynamics of
magnons in thin stripes of
the magnetic insulator yttrium-iron garnet (YIG)\cite{Cherepanov93}.
The time evolution of the momentum distribution of  
magnons in YIG  has been studied experimentally by means of 
the method of Brillouin  light scattering \cite{Demokritov06,Demidov08a,Demidov08b}. 
Note that the effective magnon Hamiltonian of YIG
is of the form (\ref{eq:hdef}) if we  
identify the operators $a_{\bd{k}}^{\dagger}$ with the
creation operators  of the relevant magnon band  in YIG \cite{Kreisel09}.
In the experimentally realized
thin stripe geometry the magnons form a quasi 
two-dimensional weakly interacting Bose gas.
If the external magnetic field $\bd{H}$ is aligned along the stripe axis
and the angle  $\theta_{\bd{k}}$
 between the magnetic field and
the magnon momentum $\bd{k}$ is not too large,
the long-wavelength energy dispersion 
of the experimentally relevant magnon band in YIG
can be approximated by \cite{Kreisel09,Kalinikos86}
 \begin{eqnarray}
  \epsilon_{\bd{k}} & =  &  [ h + \rho_{\rm ex} \bd{k}^2 + \Delta ( 1 - f_{\bd{k}} )
 \sin^2 \theta_{\bd{k}} ]^{1/2} 
 \nonumber
 \\
 &  \times &
[ h + \rho_{\rm ex} \bd{k}^2 + \Delta f_{\bd{k}} ]^{1/2}.
 \label{eq:ekyig}
 \end{eqnarray}
Here $h = \mu | \bd{H} |$ is the Zeeman  energy associated 
with the external magnetic field ($\mu$ is the magnetic moment of the localized spins 
in YIG), $\rho_{\rm ex}$
is the exchange spin-stiffness, and the energy $\Delta$ is the characteristic energy
scale associated with dipole-dipole interactions in YIG. 
To describe the experiments it is sufficient
to retain only a single magnon band and work
with an effective two-dimensional model.
The form factor $f_{\bd{k}}$ in Eq.~(\ref{eq:ekyig}) can then be approximated by \cite{Kreisel09}
 \begin{equation}
 f_{\bd{k}} \approx \frac{ 1 - e^{ - | \bd{k} | d} }{ | \bd{k} | d },
 \label{eq:form}
 \end{equation}
where $d$ is the transverse thickness of the YIG stripe.
As will be discussed in more detail in Sec.~\ref{sec:YIG},
the parameters $\rho_{\rm ex}$, $\Delta$ and $d$ entering the magnon dispersion
(\ref{eq:ekyig}) are all known from experiment, so that we may use
our model Hamiltonian (\ref{eq:hdef}) to make specific predictions of the
nonequilibrium dynamics of the magnon gas in YIG.

Note that our Hamiltonian (\ref{eq:hdef}) does not contain direct 
two-body interactions between the magnons, so that in the absence of phonons
the magnons do not interact.
A more realistic model describing the  Bose-Einstein condensation of magnons in YIG
should also include many-body interactions between the magnons\cite{Cherepanov93,Hick10}.
Although these interactions compete with the phonons to 
thermalize the magnon gas, we focus here on a regime 
where the dominant thermalization mechanism is due to magnon-phonon scattering.

\section{Rate equations from the Keldysh technique}
\label{sec:rate}

The nonequilibrium dynamics of a boson Hamiltonian of the type (\ref{eq:hdef}) 
with quadratic boson dispersion $\epsilon_{\bd{k}}$
has been 
studied in Ref.~[\onlinecite{Banyai00}] using a simple 
decoupling procedure of the equations of motion, which yields the following
rate equation for the time-dependent momentum distribution
$n_{\bd{k}} ( t ) = \langle a^{\dagger}_{\bd{k}} ( t ) a_{\bd{k}} (t ) \rangle$,
 \begin{eqnarray}
 \partial_{t} n_{\bd{k}} ( t  )  
 & = &  \frac{1}{  V} \sum_{\bd{k}^{\prime}} 
 \Bigl\{  [ 1 + n_{\bd{k}} ( t ) ]
W_{ \bd{k},     \bd{k}^{\prime} }  n_{\bd{k}^{\prime}} (t )
  \nonumber
 \\
 & & \hspace{9mm} -   [ 1 + n_{\bd{k}^{\prime}} ( t ) ] 
W_{  \bd{k}^{\prime} , \bd{k}   }  n_{\bd{k}} ( t )
 \Bigr\} .
 \label{eq:master}
 \end{eqnarray}
The transition rates are given by Fermi's golden rule\cite{Zwanzig01}
 \begin{eqnarray}
 W_{\bd{k} , \bd{k}^{\prime}} & = & 
2 \gamma^2_{ \bd{k} - \bd{k}^{\prime}} 
 b  (  \epsilon_{\bd{k}} - \epsilon_{\bd{k}^{\prime}} )
D^I_{\bd{k} - \bd{k}^{\prime}} 
 (  \epsilon_{\bd{k}} - \epsilon_{\bd{k}^{\prime}} )  ,
 \label{eq:Wk1}
 \end{eqnarray}
where  the Bose function
\begin{equation}
 b ( \omega  )  = \frac{1}{ e^{\beta \omega } -1 }
 \end{equation}
is the equilibrium distribution of the phonon bath at temperature $T = 1/ \beta$, and
 \begin{equation}
 D^I_{\bd{q}} ( \omega ) =  \pi {\rm sgn} ( \omega )
 \left[ \delta ( \omega - \omega_{\bd{q}} )  +
 \delta ( \omega + \omega_{\bd{q}} ) \right]
 \label{eq:DIdef}
 \end{equation}
is the spectral function of the symmetrized phonon propagator,
see Eq.~(\ref{eq:DIpho}) in Appendix C.
Note that for negative frequencies
the spectral function (\ref{eq:DIdef}) is negative, which is a general property of any
bosonic spectral function \cite{Negele88}.
Using the fact that $D^I_{\bd{q}} (  - \omega ) = - D^I_{\bd{q}} ( \omega )$ and
 $b ( - \omega ) = - [ 1 + b ( \omega ) ]  = - e^{ \beta \omega} b ( \omega ) $
one easily verifies that the transition rates 
satisfy detailed balance\cite{Zwanzig01}
  \begin{equation}
{W}_{  \bd{k} , \bd{k}^{\prime}  }    e^{ - \beta \epsilon_{\bd{k}^{\prime} }}  =
 {W}_{  \bd{k}^{\prime} , \bd{k} }    e^{ - \beta \epsilon_{\bd{k} }} .
 \end{equation}
Numerical solutions of Eq.~(\ref{eq:master}) for finite systems of bosons with
quadratic dispersion can be found in Ref.~[\onlinecite{Banyai00}].
However, for finite systems the Dirac $\delta$-functions in 
the transition rates (\ref{eq:Wk1})  lead to singular results.
B\'{a}nyai {\it{et al.}}\cite{Banyai00} therefore
replace the $\delta$-functions (by hand) by Lorentzians 
of finite width,
thus introducing an additional
phenomenological parameter into the problem.
This is somewhat unsatisfactory, because the final results 
can be sensitive to the width of the broadened $\delta$-functions.
Here we present an alternative solution to this problem:
within our FRG approach with hybridization cutoff 
it is straightforward to take the
physical broadening of the phonon propagators due to the coupling
to the magnon system self-consistently into account.

\begin{figure}[tb]
  \centering
\includegraphics[width=0.4\textwidth]{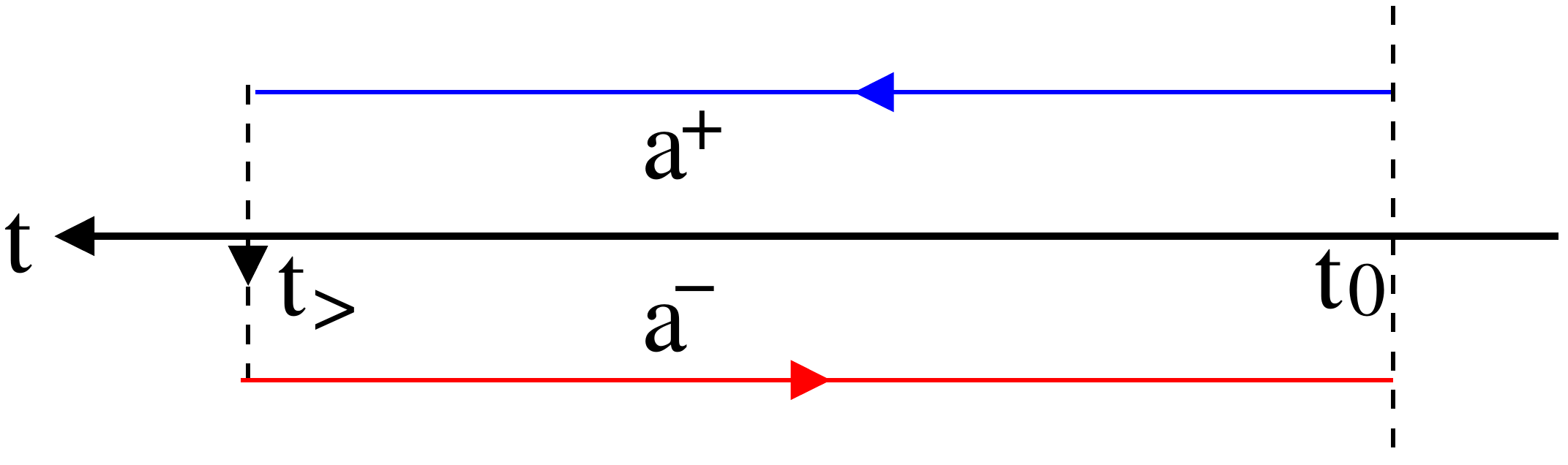}
  \caption{%
(Color online)
The Keldysh contour  connects  some initial time $t_0$
slightly above the real axis
with some final time $t_{>}$ which is larger than all times of interest.
It then crosses the real axis and goes back to
$t_0$ slightly below the real axis.
In accordance with the definition of time-ordering we draw later times to the left.
}
\label{fig:contour}
\end{figure}
\begin{figure}[tb]
  \centering
\nspace \includegraphics[width=0.45\textwidth]{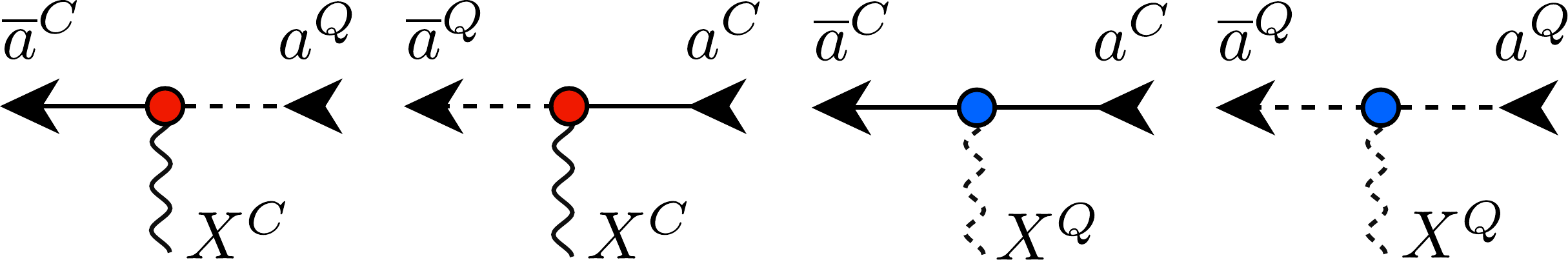}\nspace
  \caption{%
(Color online)
Graphical representation of the
bare magnon-phonon vertices
of  the 
Keldysh action (\ref{eq:Sdef}).
Solid arrows pointing into the vertices represent  $a^C$, while outgoing
solid arrows represent $\bar{a}^C$.
Dotted arrows represent the corresponding quantum components
$a^Q$ and $\bar{a}^Q$.
Wavy solid lines represent the
classical component $X^C$ of the phonon field defined in Eq.~(\ref{eq:Xdef}), while
wavy dotted lines represent the quantum component $X^Q$.
}
\label{fig:barephvertex}
\end{figure}

To obtain a FRG generalization of the  rate equation (\ref{eq:master}), an
approach based on the decoupling of the equations of motion is not useful;
instead, the nonequilibrium problem should be formulated 
in terms of self-energies which are obtained via  functional integral 
techniques.
Before developing the nonequilibrium  FRG approach, it is therefore 
instructive to carefully re-derive the rate equation (\ref{eq:master}) 
using the functional integral formulation of the 
Keldysh method\cite{Kamenev11,Kamenev04}.
Let us therefore recall that in the Keldysh method the nonequilibrium
distribution function $n_{\bd{k}} ( t )$ is related to
 the Keldysh Green function at equal times,
 \begin{equation}
 i G^K_{\bd{k}} ( t , t ) = 1 + 2 n_{\bd{k}} ( t ).
 \end{equation}
In order to calculate this, one has to consider 
$G^K_{\bd{k}} ( t , t^{\prime} )$ together with
the retarded and the advanced Green functions,
 \begin{subequations}
 \begin{eqnarray}
 i G^R_{\bd{k}} ( t , t^{\prime} ) & = &  \Theta ( t - t^{\prime} )
 \langle [ a_{\bd{k}} ( t ) , a^{\dagger}_{\bd{k}} ( t^{\prime} ) ] \rangle,
 \\
 i G^A_{\bd{k}} ( t , t^{\prime} ) & = &  -\Theta (  t^{\prime} - t )
 \langle [ a_{\bd{k}} ( t ) , a^{\dagger}_{\bd{k}} ( t^{\prime} ) ] \rangle,
 \\
i G^K_{\bd{k}} ( t , t^{\prime} ) & = & 
 \langle \{ a_{\bd{k}} ( t ) , a^{\dagger}_{\bd{k}} ( t^{\prime} ) \} \rangle,
 \end{eqnarray}
\end{subequations}
where $[ \;  ,  \; ]$ is the commutator and $\{ \; , \; \}$ is the anticommutator.
The corresponding phonon Green functions are
\begin{subequations}
 \begin{eqnarray}
 i F^R_{\bd{q}} ( t , t^{\prime} ) & = &  \Theta ( t - t^{\prime} )
 \langle [ b_{\bd{q}} ( t ) , b^{\dagger}_{\bd{q}} ( t^{\prime} ) ] \rangle,
 \\
 i F^A_{\bd{q}} ( t , t^{\prime} ) & = &  -\Theta (  t^{\prime} - t )
 \langle [ b_{\bd{q}} ( t ) , b^{\dagger}_{\bd{q}} ( t^{\prime} ) ] \rangle,
 \\
i F^K_{\bd{q}} ( t , t^{\prime} ) & = & 
 \langle \{ b_{\bd{q}} ( t ) , b^{\dagger}_{\bd{q}} ( t^{\prime} ) \} \rangle.
 \end{eqnarray}
\end{subequations}
Since the magnons in Eq.~(\ref{eq:hdef}) couple only to the
hermitian combination 
 \begin{equation}
 X_{\bd{q}} = \frac{1}{\sqrt{2}} \left( b_{\bd{q}} + b^{\dagger}_{- \bd{q}} 
 \right)
 \end{equation}
of the phonon fields,  it is convenient to define
  \begin{subequations}
 \begin{eqnarray}
 i D^R_{\bd{q}} ( t , t^{\prime} ) & = &  \Theta ( t - t^{\prime} )
 \langle [ X_{\bd{q}} ( t ) , X_{-\bd{q}} ( t^{\prime} ) ] \rangle,
 \label{eq:DR}
 \\
 i D^A_{\bd{q}} ( t , t^{\prime} ) & = &  -\Theta (  t^{\prime} - t )
 \langle [ X_{\bd{q}} ( t ) , X_{-\bd{q}} ( t^{\prime} ) ] \rangle,
 \label{eq:DA}
 \\
i D^K_{\bd{q}} ( t , t^{\prime} ) & = & 
 \langle \{ X_{\bd{q}} ( t ) , X_{-\bd{q}} ( t^{\prime} ) \} \rangle.
 \label{eq:DK}
 \end{eqnarray}
\end{subequations}
Following Kamenev\cite{Kamenev04,Kamenev11},
we represent the above Green functions as functional averages,
 \begin{subequations}
 \begin{eqnarray}
 i G^R_{\bd{k}} ( t , t^{\prime} ) & =  & 
\langle a^C_{\bd{k}} ( t ) \bar{a}^Q_{\bd{k}} ( t^{\prime} )
 \rangle \equiv i G^{CQ}_{\bd{k}} ( t , t^{\prime} ),
 \label{eq:G1}
 \\
 i G^A_{\bd{k}} ( t , t^{\prime} ) & =  & 
\langle a^Q_{\bd{k}} ( t ) \bar{a}^C_{\bd{k}} ( t^{\prime} )
 \rangle \equiv i G^{QC}_{\bd{k}} ( t , t^{\prime} ),
 \label{eq:G2}
\\
 i G^K_{\bd{k}} ( t , t^{\prime} ) & =  & 
\langle a^C_{\bd{k}} ( t ) \bar{a}^C_{\bd{k}} ( t^{\prime} )
 \rangle \equiv i G^{CC}_{\bd{k}} ( t , t^{\prime} ),
 \label{eq:G3}
\end{eqnarray}
\end{subequations}
and similarly for the phonon Green functions,
 \begin{subequations}
 \begin{eqnarray}
 i F^R_{\bd{q}} ( t , t^{\prime} ) & =  & 
\langle b^C_{\bd{q}} ( t ) \bar{b}^Q_{\bd{q}} ( t^{\prime} )
 \rangle \equiv i F^{CQ}_{\bd{q}} ( t , t^{\prime} ),
 \label{eq:F1}
 \\
 i F^A_{\bd{q}} ( t , t^{\prime} ) & =  & 
\langle b^Q_{\bd{q}} ( t ) \bar{b}^C_{\bd{q}} ( t^{\prime} )
 \rangle \equiv i F^{QC}_{\bd{q}} ( t , t^{\prime} ),
 \label{eq:F2}
\\
 i F^K_{\bd{q}} ( t , t^{\prime} ) & =  & 
\langle b^C_{\bd{q}} ( t ) \bar{b}^C_{\bd{q}} ( t^\prime )
 \rangle \equiv i F^{CC}_{\bd{q}} ( t , t^{\prime} ).
 \label{eq:F3}
\end{eqnarray}
\end{subequations}
The functional averages are defined as
 \begin{equation}
 \langle a^{\lambda}_{\bd{k}} ( t ) \bar{a}_{\bd{k}}^{\lambda^{\prime}} ( t ^{\prime} ) \rangle
 = \int {\cal{D}} [  \bar{a} , a, \bar{b}, b  ] e^{ i S [ \bar{a} , a , \bar{b} , b ] }
a^{\lambda}_{\bd{k}} ( t ) \bar{a}_{\bd{k}}^{\lambda^{\prime}} ( t ^{\prime} ),
 \end{equation}
and similarly for $\langle b^{\lambda}_{\bd{q}} ( t ) \bar{b}_{\bd{q}}^{\lambda^{\prime}} ( t ^{\prime} ) \rangle$. 
The superscripts $\lambda, \lambda^{\prime} \in \{ C, Q \}$ 
label the classical $(C)$ and quantum $(Q)$  components of the fields,
which are related to the contour components $a_{\bd{k} }^{ \pm} ( t )$ and
 $b_{\bd{k} }^{ \pm} ( t )$ via\cite{footnoteContour}
 \begin{subequations}
 \begin{eqnarray}
 a^C_{\bd{k}} ( t ) & = & \frac{1}{\sqrt{2}} \left[ 
 a_{\bd{k} }^{ + } ( t ) +   a_{\bd{k} }^{ - } ( t ) \right],
 \\
 a^Q_{\bd{k}} ( t ) & = & \frac{1}{\sqrt{2}} \left[ 
 a_{\bd{k} }^{ + } ( t ) -   a_{\bd{k} }^{ - } ( t ) \right],
 \\
 b^C_{\bd{k}} ( t ) & = & \frac{1}{\sqrt{2}} \left[ 
 b_{\bd{k} }^{ + } ( t ) +   b_{\bd{k} }^{ - } ( t ) \right],
 \\
 b^Q_{\bd{k}} ( t ) & = & \frac{1}{\sqrt{2}} \left[ 
 b_{\bd{k} }^{ + } ( t ) -   b_{\bd{k} }^{ - } ( t ) \right].
 \end{eqnarray}
 \end{subequations}
Here the superscripts $\pm$ attached to the fields
$a_{\bd{k} }^{ \pm} ( t )$ and
 $b_{\bd{k} }^{ \pm} ( t )$
refer to the upper and lower branch of the Keldysh contour shown
in Fig.~\ref{fig:contour}.
The Keldysh action
is given by
 \begin{widetext}
 \begin{eqnarray}
   S [ \bar{a} , a , \bar{b} , b ] & = & \int dt \int dt^{\prime} \Bigg\{
  \sum_{\bd{k}} ( \bar{a}_{\bd{k}}^C ( t ) , \bar{a}_{\bd{k}}^Q ( t ) )
 \left( \begin{array}{cc} 0 & (\hat{G}^A_0 )^{-1} \\
  (\hat{G}^R_0 )^{-1} & -  (\hat{G}^R_0 )^{-1} \hat{G}_0^K  (\hat{G}^A_0 )^{-1}
 \end{array} \right)_{ t t^{\prime}} 
 \left( \begin{array}{c} {a}_{\bd{k}}^C ( t^{\prime} ) \\  {a}_{\bd{k}}^Q ( t^{\prime} ) \end{array} \right)
 \nonumber
 \\ 
 & & \hspace{16mm} +\sum_{\bd{q}} ( \bar{b}_{\bd{q}}^C ( t ) , \bar{b}_{\bd{q}}^Q ( t ) )
 \left( \begin{array}{cc} 0 & (\hat{F}^A_0 )^{-1} \\
  (\hat{F}^R_0 )^{-1} & -  (\hat{F}^R_0 )^{-1} \hat{F}_0^K  (\hat{F}^A_0 )^{-1}
 \end{array} \right)_{ t t^{\prime}} 
 \left( \begin{array}{c} {b}_{\bd{q}}^C ( t^{\prime} ) \\  
 {b}_{\bd{q}}^Q ( t^{\prime} ) \end{array} \right) \Bigg\}
 \nonumber
 \\
 & - & \int dt \frac{1}{\sqrt{V}} \sum_{\bd{k} , \bd{q}} \gamma_{\bd{q}} 
 \left[ \left( \bar{a}^C_{\bd{k} + \bd{q}} a^Q_{\bd{k}} +
\bar{a}^Q_{\bd{k} + \bd{q}} a^C_{\bd{k}} \right) X^C_{\bd{q}}
+ \left( \bar{a}^C_{\bd{k} + \bd{q}} a^C_{\bd{k}} +
\bar{a}^Q_{\bd{k} + \bd{q}} a^Q_{\bd{k}} \right) X^Q_{\bd{q}}
\right].
 \label{eq:Sdef}
 \end{eqnarray}
\end{widetext}
Here we have defined the
phonon fields
 \begin{equation}
 X^{\lambda}_{\bd{q}} (t ) = \frac{1}{\sqrt{2}} \left[
 b^{\lambda}_{\bd{q}} ( t ) + \bar{b}^{\lambda}_{- \bd{q}} ( t ) \right]\; \;
, \; \; \lambda = C, Q,
  \label{eq:Xdef}
 \end{equation}
and
$\hat{G}_0^{R}, \hat{G}_0^{A}$, and $\hat{G}_0^K$ as well as
 $\hat{F}_0^{R}, \hat{F}_0^{A}$, and $\hat{F}_0^K$ are
infinite matrices in the time-labels (we omit for simplicity the momentum label),
with matrix elements given by the Green functions
(\ref{eq:G1}--\ref{eq:F3})  in the limit where the magnon-phonon coupling is switched off.
A graphical representation of the interaction vertices in Eq.~(\ref{eq:Sdef}) is  shown in
Fig.~\ref{fig:barephvertex}.
Note that the lower diagonal blocks 
of the inverse Gaussian propagators in Eq.~(\ref{eq:Sdef}) are
actually infinitesimal regularizations of the
continuous functional integral,
 \begin{subequations}
 \begin{eqnarray}
 -  (\hat{G}^R_0 )^{-1} \hat{G}_0^K  (\hat{G}^A_0 )^{-1} & = & 2 i \eta \hat{g}_0,
 \label{eq:regg}
 \\
 -  (\hat{F}^R_0 )^{-1} \hat{F}_0^K  (\hat{F}^A_0 )^{-1} & = & 2 i \eta
 \hat{f}_0,
 \label{eq:regg2}
 \end{eqnarray}
 \end{subequations}
where $\hat{g}_0$ and $\hat{f}_0$ are diagonal matrices in the time labels
which contain the magnon- and the phonon distribution functions
in the non-interacting limit,
 \begin{subequations}
 \begin{eqnarray}
 [ \hat{g}_{0} ]_{ \bd{k}, t  t^{\prime}} & = & \delta ( t - t^{\prime} ) g_{0, \bd{k}} = 
\delta ( t - t^{\prime} )
[ 1 + 2 \langle a^{\dagger}_{\bd{k}} a_{\bd{k}} \rangle_0] ,
  \label{eq:g0def}
 \\
 {[} \hat{f}_{0} ]_{\bd{q}, t t^{\prime}} & = & \delta ( t - t^{\prime} ) f_{0, \bd{q}} = 
\delta ( t - t^{\prime} )
[ 1 + 2 \langle b^{\dagger}_{\bd{q}} b_{\bd{q}} \rangle_0] .
 \hspace{10mm}
 \end{eqnarray}
 \end{subequations}

Note that we restrict ourselves to initial states without correlations. Therefore the 
initial density matrix appears only via the initial conditions for the 
independent one- and two-point Green functions \cite{Danielewicz84,Semkat00,Garny09}.
Since the Keldysh action (\ref{eq:Sdef}) is a Gaussian functional of the phonon fields,
we may perform the integration over the phonon fields exactly and obtain the following 
effective Keldysh action for the magnon fields,
 \begin{widetext}
 \begin{eqnarray}
   S [ \bar{a} , a  ] & = & \int dt \int dt^{\prime}
  \sum_{\bd{k}} ( \bar{a}_{\bd{k}}^C ( t ) , \bar{a}_{\bd{k}}^Q ( t ) )
 \left( \begin{array}{cc} 0 & (\hat{G}^A_0 )^{-1} \\
  (\hat{G}^R_0 )^{-1} & -  (\hat{G}^R_0 )^{-1} \hat{G}_0^K  (\hat{G}^A_0 )^{-1}
 \end{array} \right)_{ t t^{\prime}} 
 \left( \begin{array}{c} {a}_{\bd{k}}^C ( t^{\prime} ) \\  {a}_{\bd{k}}^Q ( t^{\prime} ) \end{array} \right)
 \nonumber
 \\
 & - & \int dt \int d t^{\prime}
\frac{1}{V} \sum_{ \bd{q}} \gamma_{\bd{q}}^2 
 \biggl\{ D^R_{0, {\bd{q}} } ( t , t^{\prime} )
 \rho^{CQ}_{ - \bd{q}} ( t )  \bigl[ \rho^{CC}_{\bd{q}} ( t^{\prime} ) +
 \rho^{QQ}_{\bd{q}} ( t^{\prime} ) \bigr]
 + D^A_{0, {\bd{q}} } ( t , t^{\prime} )
  \bigl[ \rho^{CC}_{-\bd{q}} ( t ) +
 \rho^{QQ}_{- \bd{q}} ( t ) \bigr] 
 \rho^{QC}_{  \bd{q}} ( t^{\prime} )
 \nonumber
 \\
 & & \hspace{30mm} + \frac{1}{2}  D^K_{0, {\bd{q}} } ( t , t^{\prime} )
  \bigr[ \rho^{CQ}_{-\bd{q}} ( t ) +
 \rho^{QC}_{- \bd{q}} ( t ) \bigl] \bigr[ \rho^{CQ}_{\bd{q}} ( t^{\prime} ) +
 \rho^{QC}_{\bd{q}} ( t^{\prime} ) \bigl]
\biggr\},
 \label{eq:Sadef}
 \end{eqnarray}
\end{widetext}
where $D^X_{0, \bd{q}} ( t , t^{\prime})$,  $X = R,A, K$,
denote  the non-interacting limit of the
symmetrized phonon propagators
defined in Eqs.~(\ref{eq:DR}--\ref{eq:DK}), and
we have defined
 \begin{equation}
 \rho^{\lambda \lambda^{\prime}}_{\bd{q}} ( t ) = \sum_{\bd{k}} 
 \bar{a}^{\lambda}_{\bd{k}} ( t )
 a^{\lambda^{\prime}}_{ \bd{k} + \bd{q}} ( t ),
 \; \; \; \lambda , \lambda^{\prime} \in \left\{ C, Q  \right\}.
 \end{equation}
It is convenient to collect 
the retarded ($\hat{G}^R$), advanced
($\hat{G}^A$) and Keldysh ($\hat{G}^K$) components of the
Green function into a matrix Green function,
\begin{equation}
 \mathbf{{{G}}}  =
\left(
 \begin{array}{cc}
 {[ \mathbf{{G}}]}^{CC} & {[ \mathbf{{G}}]}^{CQ} \\
 {[ \mathbf{{G}}]}^{QC} & 0
 \end{array}
 \right)  =
\left(
 \begin{array}{cc}
 \hat{G}^{K} & \hat{G}^R \\
 \hat{G}^A & 0
 \end{array}
 \right),
 \label{eq:Gmatdef}
 \end{equation}
whose inverse has the block structure
\begin{equation}
 \mathbf{G}^{-1}  = \left(
 \begin{array}{cc}
 0  & (\hat{G}^A)^{-1} \\
 (\hat{G}^R)^{-1} & -   (\hat{G}^R)^{-1}  \hat{G}^{K} (\hat{G}^A)^{-1}
 \end{array}
 \right).
 \label{eq:Gmatinv}
 \end{equation}
Defining the nonequilibrium self-energy via the matrix Dyson equation,
 \begin{equation}
 \label{eq:dyson}
 \mathbf{G}^{-1} = \mathbf{G}_0^{-1} - \mathbf{\Sigma},
 \end{equation}
we have in the Keldysh basis,
\begin{equation}
 \mathbf{\Sigma}  =
\left(
 \begin{array}{cc}
 0                    & {[\mathbf{\Sigma}]}^{CQ} \\
 {[\mathbf{\Sigma}]}^{QC} & {[\mathbf{\Sigma}]}^{QQ}
 \end{array}
 \right)  =
\left(
 \begin{array}{cc}
 0                 & \hat{\Sigma}^A \\
 \hat{\Sigma}^R & \hat{\Sigma}^K
 \end{array}
 \right) ,
 \label{eq:Sigmamatdef}
 \end{equation}
with
 \begin{equation}
 \hat{\Sigma}^K =  - [ \mathbf{G}^{-1}]^{QQ} =  ( \hat{G}^R )^{-1} \hat{G}^K ( \hat{G}^A )^{-1}.
 \label{eq:SigmaKdef}
 \end{equation}
By taking appropriate matrix elements of 
the nonequilibrium Dyson equation (\ref{eq:dyson}), we obtain a quantum kinetic equation
for the distribution function.
Actually, as reviewed in Appendix~A,
one can rewrite the kinetic equation in several different
forms, depending on the choice of basis and on the
re-shuffling of the terms in the Dyson equations.
For our purpose, it is most convenient to
parametrize the Keldysh block in terms of the distribution matrix
$\hat{g}$ by setting
\begin{equation}
 \hat{G}^K = \hat{G}^R \hat{g}^{\dagger} - \hat{g} \hat{G}^A,
 \label{eq:Gstruc}
 \end{equation}
where in the non-interacting limit the  matrix $\hat{g}$ 
reduces to the non-interacting distribution matrix $\hat{g}_0$
given in Eq.~(\ref{eq:g0def}).
It is then easy to show (see Appendix A) that the
Dyson equation (\ref{eq:dyson}) implies
 \begin{equation}
  - i ( \hat{M} \hat{g} - \hat{g}^{\dagger} \hat{M} ) = \hat{\Sigma}^{\rm in} - 
 \hat{\Sigma}^{\rm out},
 \label{eq:kineqg}
 \end{equation}
where the infinite matrices $\hat{M}$,   $\hat{\Sigma}^{\rm in}$
and  $\hat{\Sigma}^{\rm out} $ can be written as
 \begin{eqnarray}
 \hat{M} & = & \hat{M}_0 - \hat{\Sigma}^M,
 \label{eq:Mdef}
 \\
  \hat{\Sigma}^{\rm in} & = &
 i \hat{\Sigma}^{K},
 \label{eq:Sigmain}
 \\
 \hat{\Sigma}^{\rm out}  & = & 
  \frac{1}{2} 
 \left(  \hat{\Sigma}^I \hat{g}   +  \hat{g}^{\dagger}   \hat{\Sigma}^I \right).
 \label{eq:Sigmaout}
 \end{eqnarray}
Here  the matrix elements of $\hat{M}_0$ in
the momentum-time basis are
\begin{eqnarray}
  [ \hat{M}_0 ]_{ \bd{k} t , \bd{k}^{\prime} t^{\prime}}
& = & \delta_{ \bd{k} ,  \bd{k}^{\prime}  } 
 \left[   i  \delta^{\prime} ( t - t^{\prime})  - \epsilon_{\bd{k}} 
\delta ( t - t^{\prime})
  \right] ,
 \label{eq:M0def}
 \end{eqnarray}
where $\delta^{\prime} ( t ) = \frac{ d }{dt} \delta ( t ) $
is the derivative of the Dirac $\delta$-function.
Moreover, 
we have introduced
the mean (denoted by a superscript $M$) and the spectral 
(denoted by a superscript $I$ for imaginary part) 
combinations of the retarded and advanced self-energies\cite{Rammer07},
 \begin{equation}
 \hat{\Sigma}^M  =  \frac{1}{2}[ \hat{\Sigma}^R +  \hat{\Sigma}^A ]
 \quad , \quad
 \hat{\Sigma}^I  =  i [ \hat{\Sigma}^R -  \hat{\Sigma}^A ].
 \label{eq:Asymdef}
 \end{equation}
Our notation for
$\hat{\Sigma}^{\rm in}$
and  $\hat{\Sigma}^{\rm out} $ anticipates that these quantities are related
to the ``in-scattering'' and ``out-scattering'' contributions to the
collision integral of the kinetic equation.

To describe  the approach towards equilibrium,
it is useful to express our kinetic equation in terms of
Wigner transforms.  Given any infinite matrix $\hat{A}$ in the time labels with 
matrix elements $[ \hat{A}]_{ t t^{\prime} } $, we define
its Wigner transform $A (  \tau ; \omega )$ by performing a partial
Fourier transformation with respect to the time difference $s = t - t^{\prime}$,
 \begin{equation}
 A ( \tau ; \omega )
=   \int_{ - \infty}^{\infty} d s
 e^{  i \omega s } [ \hat{A}]_{  \tau + \frac{s}{2} , \tau - \frac{s}{2} },
 \label{eq:Wignerdef}
 \end{equation}
which is a function of the average time $\tau = \frac{ t + t^{\prime}}{2}$ and the
frequency $\omega$ associated with the difference $ t - t^{\prime}$.
The Wigner transform of our 
kinetic equation (\ref{eq:kineqg}) is
\begin{eqnarray}
& & 
   \partial_\tau {\rm Re} g_{\bd{k}}  ( \tau ; \omega )  
 + 2  ( \omega - \epsilon_{\bd{k}} ) {\rm Im} g_{\bd{k}} ( \tau ; \omega )
 \nonumber
\\ 
 & & \hspace{21mm}
 + i  \bigl(  \hat{\Sigma}^M \hat{g} - \hat{g}^{\dagger} \hat{\Sigma}^M  
\bigr)_{ ( \tau ; \omega )}
 \nonumber
 \\
  & = & i \Sigma^{K}_{\bd{k}} ( \tau; \omega ) -  \frac{1}{2} 
 \bigl( \hat{\Sigma}^I \hat{g} + \hat{g}^{\dagger} \hat{\Sigma}^I 
\bigr)_{ (\tau ; \omega )},
 \hspace{7mm}
 \label{eq:kingWig}
\end{eqnarray}
where we have used the fact that the Wigner transform of $\hat{g}^{\dagger}$ is
given by the function $g^{\ast} ( \tau ; \omega )$,
and $(\hat{A} \hat{B} )_{( \tau ; \omega )}$ denotes the Wigner transform 
of the product of two matrices in the time labels.
Although in general such a product does not factorize into the product 
of Wigner transforms of $\hat{A}$ and $\hat{B}$, 
there is an approximate factorization to leading order
in a gradient expansion, see Eq.~(\ref{eq:WTprod}).
Assuming that such an expansion is justified, we obtain to leading order
 \begin{eqnarray}
& & i \bigl(  \hat{\Sigma}^M \hat{g} - \hat{g}^{\dagger} \hat{\Sigma}^M  
\bigr)_{ ( \tau ; \omega )}  \approx  - 2  \Sigma^M_{\bd{k}} ( \tau ; \omega )  
 {\rm Im} g_{\bd{k}} ( \tau; \omega ),
 \hspace{7mm}
 \\
  & &
   \frac{1}{2} 
 \bigl( \hat{\Sigma}^I \hat{g} + \hat{g}^{\dagger} \hat{\Sigma}^I 
\bigr)_{ (\tau ; \omega )}    
  \approx  \Sigma^{I}_{\bd{k}} ( \tau ; \omega )
 {\rm Re} g_{\bd{k}} ( \tau; \omega ),
 \end{eqnarray}
so that our kinetic equation  (\ref{eq:kingWig}) becomes
\begin{eqnarray}
& & 
  \partial_\tau {\rm Re} g_{\bd{k}}  ( \tau ; \omega )  
 + 2  [ \omega - \epsilon_{\bd{k}}    - \Sigma^M_{\bd{k}} ( \tau ; \omega )] 
 {\rm Im} g_{\bd{k}} ( \tau ; \omega )
 \nonumber
 \\ 
 & &
 = i \Sigma^{K}_{\bd{k}} ( \tau; \omega ) -  \Sigma^{I}_{\bd{k}} ( \tau; \omega )  
 {\rm Re} g_{\bd{k}}  ( \tau ; \omega )    .
 \label{eq:kingWig2}
\end{eqnarray}
Next, let us neglect the term 
involving the imaginary part of  
$g_{\bd{k}} ( \tau ; \omega )$ and set $\omega = \epsilon_{\bd{k}}$.
Defining the quasi-particle distribution function $g_{\bd{k}} ( \tau  ) = {\rm Re} 
g_{\bd{k}} ( \tau ; \omega = \epsilon_{\bd{k}} )$ 
we arrive at the simplified kinetic equation
\begin{eqnarray}
   \partial_\tau g_{\bd{k}}  ( \tau  )  
 & = & 
 i \Sigma^{K}_{\bd{k}} ( \tau; \epsilon_{\bd{k}} ) 
 - \Sigma^{I}_{\bd{k}} ( \tau; \epsilon_{\bd{k}} )  
 g_{\bd{k}}  ( \tau ) .
 \hspace{7mm}
 \label{eq:kinqpfinal}
 \end{eqnarray}
Setting $g_{\bd{k}} ( \tau ) =  1 + 2  n_{\bd{k}} ( \tau )$,
this equation can also be written as
 \begin{equation}
 \partial_\tau 2 n_{\bd{k}}  ( \tau  )  
  =  \Sigma^{\rm in}_{\bd{k}} ( \tau ) -  \Sigma^{\rm out }_{\bd{k}} ( \tau ),
 \end{equation}
where the ``in-scattering'' and ``out-scattering'' self-energies are
given by
 \begin{subequations}
 \begin{eqnarray}
 \Sigma^{\rm in}_{\bd{k}} ( \tau ) & = &  
 i \Sigma^{K}_{\bd{k}} ( \tau; \epsilon_{\bd{k}} ),
 \label{eq:sigmainqp}
 \\
 \Sigma^{\rm out}_{\bd{k}} ( \tau ) & = &  
\Sigma^{I}_{\bd{k}} ( \tau; \epsilon_{\bd{k}} )  
[ 1 + 2 n_{\bd{k}}  ( \tau )  ].
 \label{eq:sigmaoutqp}
\end{eqnarray}
\end{subequations}
To make further progress, we need approximate expressions for the
nonequilibrium self-energies 
$\Sigma^{K}_{\bd{k}} ( \tau; \epsilon_{\bd{k}} )$ and
$\Sigma^{I}_{\bd{k}} ( \tau; \epsilon_{\bd{k}} )$. In
Appendix B we give perturbative expressions for these self-energies to second order in
the magnon-phonon coupling $\gamma_{\bd{q}}$.
Note that the simplest way to obtain the self-energies to this order is to
perform a Hartree-Fock decoupling of the
interaction term of the effective boson action (\ref{eq:Sadef}).
Within the same approximation as above (factorization of Wigner transforms, neglecting
renormalization of the magnon energies, see Appendix C for details) we obtain
 \begin{eqnarray}
 i \Sigma^K_{\bd{k}} ( \tau ; \epsilon_{\bd{k}} ) & = &
\frac{1}{  V} \sum_{\bd{k}^{\prime} } 
 \bigl\{    [ 1 + n_{\bd{k}^{\prime}} ( \tau ) ]    W_{  \bd{k}^{\prime} , \bd{k}  }
 +   W_{ \bd{k} , \bd{k}^{\prime} } n_{\bd{k}^{\prime}} ( \tau )     
 \bigr\} ,
 \label{eq:SigmaKpert}
 \nonumber
 \\
 & &
 \\
 \Sigma^I_{\bd{k}} ( \tau ; \epsilon_{\bd{k}} ) & = &
\frac{1}{  V} \sum_{\bd{k}^{\prime}} 
 \bigl\{  [ 1 + n_{\bd{k}^{\prime}} ( \tau ) ]    W_{ \bd{k}^{\prime}  , \bd{k} }
 -   W_{ \bd{k} ,  \bd{k}^{\prime} } n_{\bd{k}^{\prime}} ( \tau )     
 \bigr\} ,
 \nonumber
 \\
 & &
 \label{eq:SigmaIpert}
 \end{eqnarray}
where the transition rates $W_{\bd{k} , \bd{k}^{\prime} }$ are given in
Eq.~(\ref{eq:Wk1}).
Substituting Eqs.~(\ref{eq:SigmaKpert}, \ref{eq:SigmaIpert})
into the kinetic equation (\ref{eq:kinqpfinal}) 
and renaming $\tau \rightarrow t$ 
we finally arrive at the rate equation (\ref{eq:master}).

\begin{figure}[tb]
  \centering
\includegraphics[width=0.4\textwidth]{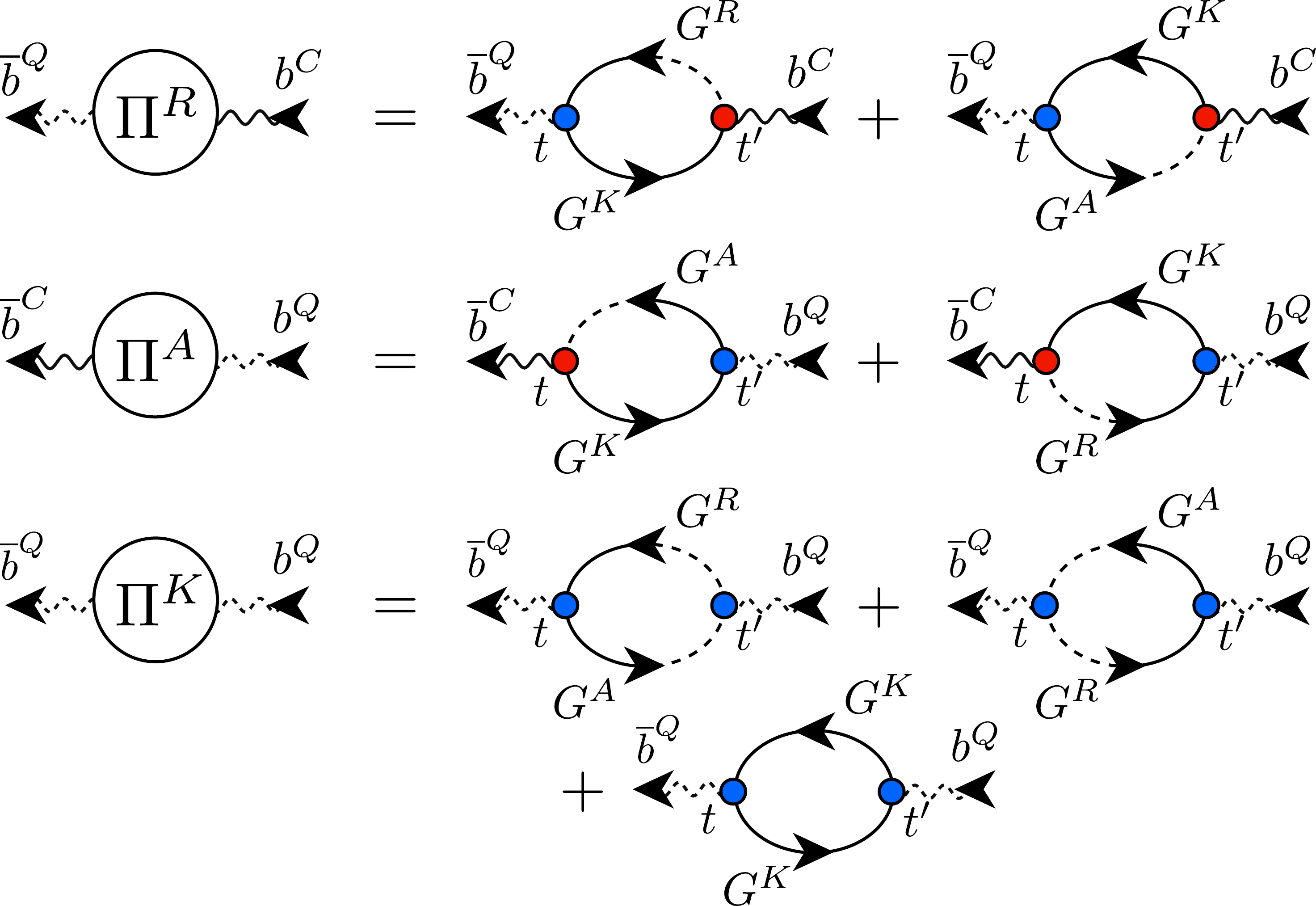}\nspace
\vspace{5mm}
  \vspace{-4mm}
  \caption{%
(Color online)
Diagrams representing the
phonon self-energies $\Pi^R$, $\Pi^A$ and $\Pi^K$ 
to second order in the magnon-phonon interaction.
Wavy lines represent the external phonon fields $b^C$ and $b^Q$, while arrows represent
the magnon Green functions $G^R$, $G^A$ and $G^K$.
We use the convention that the classical field components are represented by
solid lines, while the quantum components are represented by dotted lines,
see also the caption of Fig.~\ref{fig:barephvertex}.
If we introduce in our FRG approach  
a hybridization cutoff only in the phonon propagators, these diagrams are
finite at the initial cutoff scale.
}
\label{fig:selfpho}
\end{figure}

For finite systems the $\delta$-functions
in the phonon spectral function (\ref{eq:DIdef})
appearing in the perturbative expression (\ref{eq:Wk1})
for the transition rates should be regularized. 
The simplest possibility is to perform an ad hoc replacement
of  the $\delta$-functions by Lorentzians with finite width \cite{Banyai00},
but this procedure   
introduces an additional phenomenological parameter into the problem.
Microscopically, the coupling of the phonons to the magnon system
automatically  generates such a broadening.
We therefore introduce renormalized phonon propagators
involving the
phonon self-energies $\Pi^{X}_{\bd{q}} ( \tau ; \omega )$, where
$X =R,A,K$.
Defining the mean and the spectral combinations of the retarded and advanced phonon self-energies,
\begin{equation}
\hat{\Pi}^M  =  \frac{1}{2}[ \hat{\Pi}^R +  \hat{\Pi}^A ]
\quad , \quad
\hat{\Pi}^I  =  i [ \hat{\Pi}^R -  \hat{\Pi}^A ],
\label{eq:Asymdefphonon}
\end{equation}
the Wigner transforms of the renormalized
phonon propagators can be written as
 \begin{subequations}
\begin{eqnarray}
F^R_{\bd{q}} (\tau; \omega ) & = & \frac{1}{\omega - \omega_{\bd{q}} - 
\Pi^R_{\bm{q}}(\tau; \omega) },
\\
F^A_{\bd{q}} (\tau; \omega ) & = & \frac{1}{\omega - \omega_{\bd{q}} - 
\Pi^A_{\bm{q}}(\tau; \omega) },
\\
F^I_{\bd{q}} (\tau; \omega ) & = &
 \frac{ \Pi^I_{\bm{q}}(\tau; \omega) }{ \left[  \omega - \omega_{\bd{q}} 
 - \Pi^M_{\bm{q}}(\tau; \omega) \right]^2 + \left[   \frac{1}{2}\Pi^I_{\bm{q}}(\tau; \omega)  \right]^2 },
 \nonumber
\\
& & 
 \\
i F^K_{\bd{q}} (\tau; \omega) & = & 
\coth \left( \frac{\beta \omega }{ 2} \right)  F^I_{\bd{q} } (\tau ; \omega).
\hspace{9mm}
\end{eqnarray}
 \end{subequations}
The diagrams of the phonon self-energies up to second order in the magnon-phonon 
coupling $\gamma_{\bm{q}}$ are shown in Fig.~\ref{fig:selfpho}.
Approximating the Wigner transforms as discussed above, and neglecting the 
real parts of the self-energies, we 
have $\Pi^M \approx 0$ and $\Pi^R  \approx \frac{i}{2} \Pi^I \approx - \Pi^A
$, where the spectral component of the phonon self-energy is
\begin{eqnarray}
\Pi_{\bm{q}}^{\mathrm{I}} \left(\tau; \omega\right) & = &
\frac{2\pi}{V} \gamma_{\bm{q}}^{2} \sum_{\bm{k}} \delta \left( 
\omega + \epsilon_{\bm{k}-\bm{q}} - \epsilon_{\bm{k}} \right)
\nonumber
\\
& & \times  \left\{ 
n_{\bm{k}-\bm{q}} \left(\tau\right) - n_{\bm{k}} \left( \tau \right) \right\} .
 \label{eq:PiI}
\end{eqnarray}
To obtain regular results also for finite magnon gases,
we evaluate the right-hand side of Eq.~(\ref{eq:PiI})
in the limit of infinite volume, where
$\Pi_{\bm{q}}^{\mathrm{I}} \left(\tau; \omega\right)$
is a continuous function of all of its arguments. 
Let us emphasize that we take the limit $V \rightarrow \infty$ 
only in the phonon self-energies, so that with the resulting regular
transition rates we can still use our rate equation  (\ref{eq:master})  
to describe the kinetics of a finite magnon system with discrete spectrum.
\\
\\

\section{Functional renormalization
group approach}
\label{sec:FRG}

Formally exact FRG flow equations for the irreducible vertices of  any
bosonic many-body systems out of equilibrium\cite{Gasenzer08,Kloss11}
can be obtained as a special case of the general hierarchy of FRG flow
equations\cite{Kopietz10} which follows from the Wetterich equation\cite{Wetterich93}.
While it is straightforward to write down these equations,
the proper identification of a sensible cutoff procedure which does
not violate causality and respects the fluctuation-dissipation theorem
under equilibrium conditions is rather delicate problem.
It turns out that for the coupled magnon-phonon system considered 
here a modified hybridization cutoff scheme \cite{Jacobs10,Kloss11}
is most convenient. In this scheme the infinitesimal imaginary parts $ \pm i \eta$
defining the boundary conditions of the retarded and advanced propagators
are replaced by finite quantities $\pm i \Lambda$,
where $\Lambda$ is identified with the flow-parameter of the renormalization
group procedure. 
Moreover, in the hybridization cutoff scheme the same replacement
 is also made in the 
Keldysh blocks of the inverse Gaussian propagators, see Eqs.~(\ref{eq:regg}) and
(\ref{eq:regg2}).  
For bosonic systems, however, one should be careful to take into account
that the spectral function should be positive for positive frequencies, and
negative for negative frequencies \cite{Negele88}. 
The imaginary part of the spectral component of the self-energy
therefore changes sign at zero frequency.
Hence, in the bosonic version of the  hybridization cutoff 
scheme the infinitesimal imaginary part $\eta$ should be replaced by
$ \eta \rightarrow \Lambda {\rm sgn} ( \omega )$.
In contrast, for fermions the spectral function is positive for all
frequencies, so that the hybridization cutoff should be
introduced by replacing $\eta \rightarrow \Lambda$.

In this work,  we introduce a hybridization cutoff {\it{only
in the phonon propagators}}, while keeping the 
$\pm i \eta$ in the magnon propagators infinitesimal. In this way we do not
artificially smear out the magnon distribution, so that
we can still describe the nonequilibrium dynamics in terms of a kinetic
equation describing the evolution of the occupation probabilities
of sharp energy levels. 
The cutoff-dependent Keldysh action
$S_{\Lambda} [ \bar{a}, a , \bar{b} , b ]$ with a hybridization cutoff in the phonon
propagators can 
be obtained from the action $S [ \bar{a} , a , \bar{b} , b ]$
given in Eq.~(\ref{eq:Sdef}) by replacing 
the Gaussian phonon propagators by cutoff-dependent propagators,
 \begin{widetext}
 \begin{eqnarray}
   S_{\Lambda} [ \bar{a} , a , \bar{b} , b ] & = & \int dt \int dt^{\prime} \Bigg\{
  \sum_{\bd{k}} ( \bar{a}_{\bd{k}}^C ( t ) , \bar{a}_{\bd{k}}^Q ( t ) )
 \left( \begin{array}{cc} 0 & (\hat{G}^A_0 )^{-1} \\
  (\hat{G}^R_0 )^{-1} & 2 i \eta \hat{g}_0
 \end{array} \right)_{ t t^{\prime}} 
 \left( \begin{array}{c} {a}_{\bd{k}}^C ( t^{\prime} ) \\  {a}_{\bd{k}}^Q ( t^{\prime} ) \end{array} \right)
 \nonumber
 \\ 
 & & \hspace{16mm} +\sum_{\bd{q}} ( \bar{b}_{\bd{q}}^C ( t ) , \bar{b}_{\bd{q}}^Q ( t ) )
 \left( \begin{array}{cc} 0 & (\hat{F}^A_{ \Lambda} )^{-1} \\
  (\hat{F}^R_{\Lambda} )^{-1} & -  (\hat{F}^R_{ \Lambda} )^{-1} \hat{F}_{ \Lambda}^K  (\hat{F}^A_{\Lambda} )^{-1}
 \end{array} \right)_{ t t^{\prime}} 
 \left( \begin{array}{c} {b}_{\bd{q}}^C ( t^{\prime} ) \\  
 {b}_{\bd{q}}^Q ( t^{\prime} ) \end{array} \right) \Bigg\}
 \nonumber
 \\
 & - & \int dt \frac{1}{\sqrt{V}} \sum_{\bd{k} , \bd{q}} \gamma_{\bd{q}} 
 \left[ \left( \bar{a}^C_{\bd{k} + \bd{q}} a^Q_{\bd{k}} +
\bar{a}^Q_{\bd{k} + \bd{q}} a^C_{\bd{k}} \right) X^C_{\bd{q}}
+ \left( \bar{a}^C_{\bd{k} + \bd{q}} a^C_{\bd{k}} +
\bar{a}^Q_{\bd{k} + \bd{q}} a^Q_{\bd{k}} \right) X^Q_{\bd{q}}
\right].
 \label{eq:SLambdadef}
 \end{eqnarray}
\end{widetext}
Because for $\Lambda \rightarrow \infty$ all phonon propagators vanish,
the FRG flow should be integrated with the boundary condition that
all irreducible pure magnon vertices vanish for $\Lambda \rightarrow \infty$.
This implies that all closed loops involving only magnon propagators 
do not vanish in this limit, leading to a non-trivial
initial condition for the FRG flow.
This is analogous to the bosonic momentum transfer cutoff 
scheme \cite{Schuetz05,Kopietz10}  in coupled fermion-boson systems,
where at the initial scale all pure boson vertices are finite.
In particular, in the present problem the phonon self-energy diagrams
shown in Fig.~\ref{fig:selfpho} are finite at the initial scale.
Neglecting the renormalization of the real part
of the phonon self-energies, the scale-dependent 
phonon propagators are in our cutoff scheme given by
\begin{subequations}
 \begin{eqnarray}
F^R_{\Lambda, \bd{q}} (\tau; \omega) & = & \frac{1}{\omega - \omega_{\bd{q}} 
+ i \Lambda \mathrm{sgn} \omega
+ \frac{i}{2} \Pi^I_{\Lambda, \bm{q}}(\tau; \omega)
 },
\\
F^A_{\Lambda , \bd{q}} (\tau; \omega) & = & \frac{1}{\omega - \omega_{\bd{q}} 
 - i \Lambda \mathrm{sgn} \omega  - \frac{i}{2} \Pi^I_{\Lambda, \bm{q}}(\tau ;  \omega)  },
\\
F^I_{ \Lambda, \bd{q}} (\tau; \omega) & = &
\frac{ 2 \Lambda \mathrm{sgn} \omega  + \Pi^I_{\Lambda, \bm{q}}(\tau ;  \omega)}{ (  \omega - \omega_{\bd{q}} )^2 + 
 \bigl[  \Lambda {\rm sgn} \omega +  \frac{1}{2}\Pi^I_{\Lambda , \bm{q}}(\tau; \omega)  
 \bigr]^2 } ,
 \nonumber
    \\
 & &
\\
i F^K_{   \Lambda  , \bd{q}} (\tau; \omega) & = & 
\coth \left( \frac{\beta \omega }{ 2} \right)  
F^I_{ \Lambda , \bd{q} } (\tau ; \omega),
 \hspace{9mm}
 \end{eqnarray}
\end{subequations}
where the imaginary part of the scale-dependent phonon self-energy
is approximated by an expression similar to Eq.~(\ref{eq:PiI}),
 \begin{eqnarray}
 \Pi_{\Lambda , \bm{q}}^{\mathrm{I}} \left(\tau; \omega\right) & = &
 \frac{2\pi}{V} \gamma_{\bm{q}}^{2} \sum_{\bd{k}} \delta \left( 
\omega + \epsilon_{\bm{k}-\bm{q}} - \epsilon_{\bm{k}} \right)
\nonumber
\\
& & \times  \left\{ 
n_{\Lambda , \bm{k}-\bm{q}} \left(\tau\right) - n_{\Lambda, 
\bm{k}} \left( \tau \right) \right\} ,
 \label{eq:PiI2}
\end{eqnarray}
where $n_{\Lambda , \bd{k}} ( \tau )$ is the scale-dependent magnon distribution function,
and it is understood that for our numerical calculations we replace
$V^{-1} \sum_{\bd{k}} \rightarrow \int d^2 k /(2 \pi )^2$
to obtain a smooth phonon self-energy in our two-dimensional system.
 Note that the scale parameter $\Lambda$ can be considered as
an additional contribution to the imaginary part of the 
phonon self-energy, which arises from the hybridization 
between phonons and  magnons.
Introducing the hybridization function
\begin{equation}
 \Lambda_{\bm{q}}(\tau;  \omega) \equiv \Lambda  {\rm{sgn}} \omega
+    \frac{1 }{2}  \Pi^I_{\Lambda , \bm{q} }(\tau;  \omega),
 \label{eq:hyb}
\end{equation}
the  symmetrized phonon propagators can be written as
\begin{subequations}
 \begin{eqnarray}
 D^R_{ \Lambda , \bd{q}} ( \omega ) &  = &
\frac{ \omega_{\bd{q}}    -  i {\Lambda}_{\bm{q}}(\tau; \omega)    }{ 
   \omega^2 - \left[ \omega_{\bd{q}} -  i {\Lambda}_{\bm{q}}(\tau; \omega) 
\right]^2    },\qquad
 \label{eq:DRlam}
 \\
 D^A_{\Lambda , \bd{q}} ( \omega ) & = &  
\frac{ \omega_{\bd{q}}    +  i {\Lambda}_{\bm{q}}(\tau; \omega)   }{ 
 \omega^2 - \left[ 
\omega_{\bd{q}} +  i  {\Lambda}_{\bm{q}}(\tau; \omega) 
\right]^2  },\qquad
\label{eq:DAlam}
 \\
 D^I_{\Lambda, \bd{q}} ( \omega ) & = & 
 \frac{  {\Lambda}_{\bm{q}}(\tau ;  \omega) }{ (  \omega - \omega_{\bd{q}} )^2
 + {\Lambda}^2_{\bm{q}}(\tau ;  \omega) }
 \nonumber
 \\
 & + &  \frac{ {\Lambda}_{\bm{q}}(\tau ;  \omega) }{ (  \omega + \omega_{\bd{q}} )^2
 + {\Lambda}^2_{\bm{q}}(\tau;  \omega) },
 \label{eq:DILambdapho}
 \\
 i D^K_{ \Lambda , \bd{q}} ( \omega ) & = &   \coth \left( \frac{\beta \omega }{ 2} \right)
    D^I_{\Lambda, \bd{q}} ( \omega ).
 \label{eq:DKLambdapho}
 \end{eqnarray}
\end{subequations}

\begin{figure}[tb]
  \centering
\includegraphics[width=0.4\textwidth]{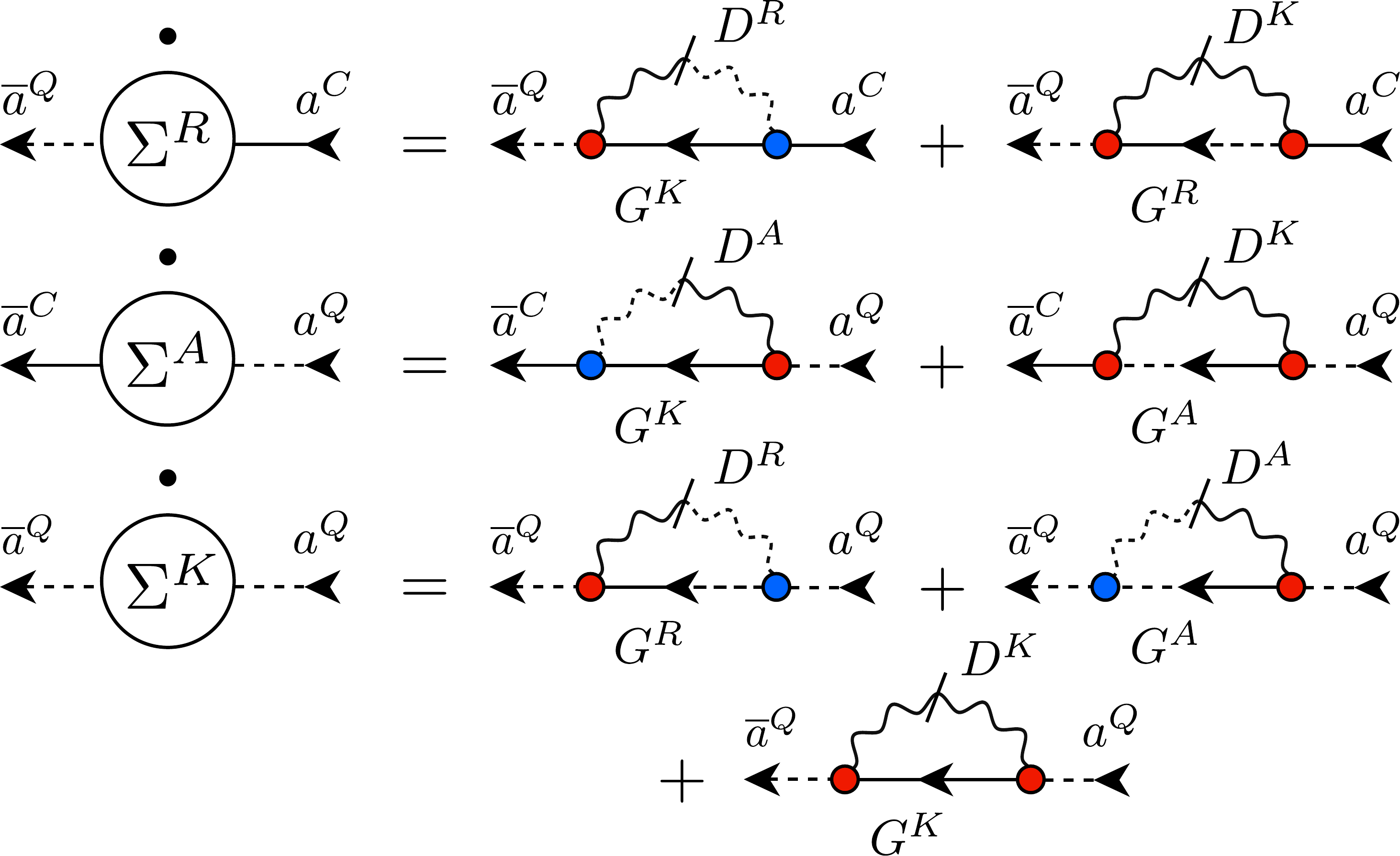}\nspace
\vspace{5mm}
  \vspace{-4mm}
  \caption{%
(Color online)
Graphical representation of the truncated FRG flow equations for the nonequilibrium
magnon self-energies 
with hybridization cutoff in the phonon propagators.
The dots over the self-energy symbols on the left-hand side
represent the derivative with respect to the 
cutoff parameter $\Lambda$. Slashed wavy lines on the right-hand side
represent single-scale phonon propagators,
which can be obtained by taking the partial derivative
of Eqs.~(\ref{eq:DRlam}--\ref{eq:DKLambdapho}) with respect to
the explicit  $\Lambda$-dependence.
Note that the diagrams on the right-hand side can be obtained
from the self-energy diagrams
given in Appendix B (see Fig.~\ref{fig:diagramsself}) by simply
replacing the phonon-propagators by  the corresponding single-scale propagators.
}
\label{fig:flowself}
\end{figure}

The general form of nonequilibrium FRG flow equations for
interacting bosons in the Keldysh basis has been written down
in Refs.~[\onlinecite{Gasenzer08,Kloss11}].
Here we use a simple truncation where the renormalization of the magnon-phonon
vertex is neglected.
The relevant FRG flow equations for the nonequilibrium magnon self-energies
with hybridization cutoff in the phonon propagators are shown graphically
in Fig.~\ref{fig:flowself}.
For the explicit evaluation of these expressions, we 
use the same approximations as in the derivation of the
perturbative rate equation (\ref{eq:master}) outlined
in Sec.~\ref{sec:rate}, namely:
 \begin{enumerate}
 \item Keep only the leading terms in the gradient expansion so that
  Wigner transforms of products of infinite matrices in the time labels factorize.

\item Neglect the contributions in the kinetic equation describing the
renormalization of the magnon energies.

\item Take  the frequency argument of the self-energies on resonance,
$\omega = \epsilon_{\bd{k}}$.

\end{enumerate}
After straightforward manipulations analogous to those 
described in Sec.~\ref{sec:rate} we obtain a system of
three coupled partial differential equations
for the scale-dependent magnon distribution function
$n_{\Lambda, \bd{k}} ( \tau )$ and the
scale-dependent nonequilibrium self-energies
 $
 \Sigma^{K}_{ \Lambda, \bd{k}} ( \tau ; \epsilon_{\bd{k}} )$ and 
$
 \Sigma^{I}_{ \Lambda, \bd{k}} ( \tau ;  \epsilon_{\bd{k}} )$.
The FRG version of the kinetic equation 
is formally identical with the perturbative kinetic equation (\ref{eq:kinqpfinal}),
 \begin{equation}
    \partial_\tau 2 n_{\Lambda , \bd{k}}  ( \tau  )  
 =   i \Sigma^{K}_{ \Lambda, \bd{k}} ( \tau ; \epsilon_{\bd{k}} ) -  
\Sigma^{I}_{\Lambda , \bd{k}} ( \tau ; \epsilon_{\bd{k}} )  
\left[1 +  2 n_{\Lambda , \bd{k}}  ( \tau ) \right]   .
 \label{eq:kinmaster}
\end{equation}
In contrast to the perturbative equation~(\ref{eq:kinqpfinal})
where the nonequilibrium self-energies can be expressed as integrals
depending on the distribution function \mbox{[see Eqs.~(\ref{eq:SigmaKpert}, 
\ref{eq:SigmaIpert})]}, in our FRG approach the
self-energies are determined by flow equations relating their derivatives
with respect to the cutoff $\Lambda$ to the $\Lambda$-dependent distribution function,
 \begin{eqnarray}
\partial_{\Lambda} i \Sigma^K_{\Lambda , \bd{k}} ( \tau ; \epsilon_{\bd{k}} ) & = &
\frac{1}{  V} \sum_{\bd{k}^{\prime} } 
 \bigl\{    [ 1 + n_{\Lambda, \bd{k}^{\prime}} ( \tau ) ]     
 \dot{W}_{  \bd{k}^{\prime} , \bd{k}  } 
 \nonumber
 \\
 & &
 \hspace{9mm} 
 +   \dot{W}_{ \bd{k} , \bd{k}^{\prime} } n_{\Lambda, \bd{k}^{\prime}} ( \tau )     
 \bigr\} ,
 \label{eq:SigmaKFRG}
 \\
 \partial_{\Lambda} \Sigma^I_{\Lambda , \bd{k}} ( \tau ; \epsilon_{\bd{k}} ) & = &
\frac{1}{  V} \sum_{\bd{k}^{\prime}} 
 \bigl\{  [ 1 + n_{\Lambda, \bd{k}^{\prime}} ( \tau ) ]    \dot{W}_{ \bd{k}^{\prime}  , \bd{k} }
 \nonumber
 \\
 & & \hspace{9mm}
-   \dot{W}_{ \bd{k} ,  \bd{k}^{\prime} } n_{\Lambda, \bd{k}^{\prime}} ( \tau )     
 \bigr\} .
 \label{eq:SigmaIFRG}
 \end{eqnarray}
Here  $\dot{W}_{\bd{k} , \bd{k}^{\prime} }$ 
is the partial derivative of the cutoff-dependent transition rate 
with respect to the explicit $\Lambda$-dependence,
 \begin{eqnarray}
 \dot{W}_{\bd{k} ,\bd{k}^{\prime}}  & = &   
 2  \gamma_{\bd{k} - \bd{k}^{\prime}}^2 \dot{D}^I_{\Lambda , \bd{k} - \bd{k}^{\prime}} ( 
\epsilon_{\bd{k}} - \epsilon_{\bd{k}^{\prime}} )
b (      \epsilon_{\bd{k}} -  \epsilon_{\bd{k}^{\prime}}   ) ,
 \end{eqnarray}
which has the same structure as Eq.~(\ref{eq:Wk1}), but with
the phonon spectral function $ {D}^I_{\Lambda, \bd{q}} ( \omega )$
replaced by 
 \begin{eqnarray}
 \dot{D}^I_{\Lambda, \bd{q}} ( \omega )  
 &  = &   
 \pi \mathrm{sgn} \omega [  
L_{ {\Lambda}_{\bm{q}}(\tau ;  \omega)}^{\prime} ( \omega - \omega_{\bd{q}} )
 +  L_{ {\Lambda}_{\bm{q}}(\tau ;  \omega)}^{\prime} ( \omega + \omega_{\bd{q}} ) ],
 \nonumber
 \\
 & &
 \label{eq:Ddottau}
 \end{eqnarray}
where
 \begin{equation}
 L^{\prime}_{\Lambda} ( \omega ) = 
\partial_{\Lambda}  \left[
\frac{1}{\pi} \frac{ \Lambda}{ \omega^2 + \Lambda^2}
 \right]
  =   \frac{1}{\pi}
 \frac{ \omega^2 - \Lambda^2 }{ ( \omega^2 + \Lambda^2)^2 }
 \end{equation}
is the derivative of a normalized Lorentzian with respect to its width.
Note that the right-hand side of Eq.~(\ref{eq:Ddottau}) implicitly depends
on the time $\tau$ via the $\tau$-dependence of the phonon self-energy
$\Pi^I_{\Lambda , \bd{q}} ( \tau ; \omega )$ given
in Eq.~(\ref{eq:PiI}).
To simplify the numerical evaluation of the above system of integro-differential equations,
we shall replace in Sec.~\ref{sec:YIG} the phonon self-energy
by its equilibrium value at vanishing cutoff,
which is approached for $\tau \rightarrow \infty$.
In other words, we replace
 \begin{equation}
 \Pi^I_{\Lambda, \bd{q}} ( \tau ; \omega ) \rightarrow
  \Pi^I_{\Lambda=0, \bd{q}} ( \tau = \infty ; \omega ).
 \end{equation}
This procedure can be justified by noting that
for sufficiently large values of the cutoff parameter $\Lambda$
the contribution from the phonon self-energy
to the hybridization function
$\Lambda_{\bm{q}}(\tau;  \omega)$ defined in Eq.~(\ref{eq:hyb})
can be neglected, so that the
only for small $\Lambda$ the effect of the phonon self-energy
becomes important; but in this regime the
magnon distribution is already close to its equilibrium value.

Unfortunately, the above system of equations violates particle number conservation, in 
contrast to the perturbative kinetic equation (\ref{eq:master}),
which obviously implies
 \begin{equation}
 \frac{ \partial}{\partial t} \sum_{\bd{k}} n_{\bd{k}} ( t ) =0.
 \end{equation}
The fact that Eqs.~(\ref{eq:kinmaster}--\ref{eq:SigmaIFRG})
violate particle number conservation
is related to our approximation of 
neglecting the renormalization of the real part of the magnon energies.
Note that particle number conservation 
is related to the $U(1)$-symmetry of the bare action 
$ S_{\Lambda} [ \bar{a} , a , \bar{b} , b ]$
given in Eq.~(\ref{eq:SLambdadef}).
To see this more clearly,
let us multiply the magnon fields in 
Eq.~(\ref{eq:SLambdadef}) by an arbitrary 
scale- and time-dependent phase factor $(X = C,Q)$,
 \begin{equation}
 a_{\bd{k}}^X ( t )  =  e^{ i \alpha_{\Lambda} ( t ) }  
 a_{\bd{k}}^{\prime X} ( t ) \; \; ,  \; \; 
 \bar{a}_{\bd{k}}^X ( t )  =  e^{ - i \alpha_{\Lambda} ( t ) }  
 \bar{a}_{\bd{k}}^{\prime X} ( t ).
 \end{equation}
Then the Keldysh action (\ref{eq:SLambdadef})  changes as follows,
 \begin{eqnarray}
 & &   S_{\Lambda} [ e^{ - i \alpha} \bar{a}^{\prime} , e^{ i \alpha} a^{\prime} , \bar{b} , b ]
 =  S_{\Lambda} [ \bar{a}^{\prime} , a^{\prime} , \bar{b} , b ]
 \nonumber
 \\
 & &- \int dt \frac{ d \alpha_{\Lambda} ( t )}{dt} 
 \sum_{\bd{k}}
 \left[ \bar{a}^{\prime C}_{\bd{k}} ( t ) a_{\bd{k}}^{\prime Q } ( t ) + 
 \bar{a}^{\prime Q}_{\bd{k}} ( t ) a_{\bd{k}}^{\prime C } ( t )
 \right].
 \hspace{7mm}
 \end{eqnarray}
The additional term can be absorbed via a shift in the
retarded and advanced self-energies,
 \begin{subequations}
 \begin{eqnarray}
{ [} \hat{\Sigma}^R ]_{ t t^{\prime}}  & \rightarrow &
  [ \hat{\Sigma}^R ]_{ t t^{\prime}}   + \delta ( t - t^{\prime} ) 
\frac{ d \alpha_{\Lambda} ( t )}{dt} ,
 \\
 {[} \hat{\Sigma}^A ]_{ t t^{\prime}}  & \rightarrow &
  [ \hat{\Sigma}^A ]_{ t t^{\prime}}   + \delta ( t - t^{\prime} ) 
\frac{ d \alpha_{\Lambda} ( t )}{dt} .
 \end{eqnarray}
\end{subequations}
The extra terms drop out in 
the spectral self-energy \mbox{$\hat{\Sigma}^I = i [ \hat{\Sigma}^R - \hat{\Sigma}^A ]$}, 
but survive in the
average  \mbox{$\hat{\Sigma}^M = \frac{1}{2} [ \hat{\Sigma}^R + \hat{\Sigma}^A ]$}. 
By completely neglecting the symmetric combination $\hat{\Sigma}^M$
in our derivation of Eq.~(\ref{eq:kinmaster}),
we have lost the possibility of explicitly implementing the
condition of particle number conservation.
To impose the constraint of constant particle number on the FRG flow, we 
add the corresponding symmetric self-energy $\hat{\Sigma}^M$
to the left-hand side of the kinetic equation.
For our purpose, it is sufficient to approximate
 \begin{equation}
 [ \hat{\Sigma}^M ]_{ t  t^{\prime}} = - \delta ( t - t^{\prime} ) \mu_{\Lambda} ( t ).
 \end{equation}
The only effect of $\mu_{\Lambda} ( t )$ on the kinetic equation (\ref{eq:kinmaster})
is to replace the derivative
$\partial_{\tau}$ by the corresponding covariant derivative 
 $ \partial_{\tau}
 + \mu_{\Lambda} ( \tau )$.
Note that the time-dependent Lagrange multiplier 
 $\mu_{\Lambda} ( \tau )$
enforcing the constraint of particle-number conservation
appears in the quantum dynamics as a gauge-field~\cite{Jolicoeur91}.
As a result, Eq.~(\ref{eq:kinmaster}) should be replaced by
 \begin{eqnarray}
 & &
    [ \partial_\tau   + \mu_{\Lambda} ( \tau ) ]   2 n_{\Lambda , \bd{k}}  ( \tau  )  
 \nonumber
 \\
 & = &
   i \Sigma^{K}_{ \Lambda, \bd{k}} ( \tau ; \epsilon_{\bd{k}} ) 
   -  
\Sigma^{I}_{\Lambda , \bd{k}} ( \tau ; \epsilon_{\bd{k}} )  
\left[ 1 + 2 n_{\Lambda , \bd{k}}  ( \tau )  \right]   .
 \label{eq:kinmastermu}
\end{eqnarray}
The energy $\mu_{\Lambda} ( \tau )$ plays the role of a
scale- and time-dependent Lagrange multiplier, which  should be adjusted such that
the total particle number $N =  \sum_{\bd{k}} n_{\Lambda , \bd{k}} ( \tau )$ is
conserved for all times $\tau$ and for all values of the cutoff $\Lambda$, implying
 \begin{equation}
 \partial_{\tau} N = \sum_{\bd{k}} \partial_{\tau} n_{\Lambda , \bd{k}} ( \tau ) =0.
 \end{equation}
Substituting Eq.~(\ref{eq:kinmastermu})  into the right-hand side of
this equation, we conclude that
the  scale- and time dependent  Lagrange multiplier is explicitly given by
 \begin{eqnarray}
 \mu_{\Lambda} ( \tau ) & = &  \frac{1}{2N}
 \sum_{\bd{k}}   
 \Bigl\{  i \Sigma^{K}_{ \Lambda, \bd{k}} ( \tau ; \epsilon_{\bd{k}} ) 
 \nonumber
 \\ 
 &  &  \hspace{7mm} - 
\Sigma^{I}_{\Lambda , \bd{k}} ( \tau ; \epsilon_{\bd{k}} )  
\left[1 +  2 n_{\Lambda , \bd{k}}  ( \tau )  \right] \Bigr\}.
 \label{eq:mures}
 \end{eqnarray}
Combining
Eqs.~(\ref{eq:kinmastermu}) and (\ref{eq:mures})
with Eqs.~({\ref{eq:SigmaKFRG}), (\ref{eq:SigmaIFRG}) and (\ref{eq:PiI}) 
we obtain a closed system of particle-number conserving FRG flow equations 
which can be used to calculate the nonequilibrium time evolution of the magnon 
distribution function. In the following section we shall use these 
equations to calculate the thermalization of the magnon gas in YIG.

To conclude this section, let us summarize and justify our main approximations. 
First of all, we have retained only the leading terms in the gradient expansion, which
amounts to the factorization of Wigner transforms. This procedure is justified 
if all functions vary sufficiently slowly in space and time. For YIG this 
approximation seems to be well justified, because an alternative description in 
terms of the phenomenological Landau-Liftshitz equation (which rely on similar assumptions) has been highly successful\cite{Kalinikos86}. 
Moreover, we have set the energy argument of all functions
on resonance. Also this approximation is justified for YIG, because 
experimentally magnons in YIG are known to behave as well-defined quasiparticles.
Finally, we have neglected the renormalization of the magnon energies. By fitting
the magnon dispersion in our final expressions to experimental data, we implicitly take this effect
into account.

\section{Thermalization of magnons in YIG}
\label{sec:YIG}

Quantized spin-waves (magnons) in ordered magnets
obey to a very good approximation Bose statistics, so that
magnetic insulators can serve as relatively easily accessible model systems for
investigating  the nonequilibrium dynamics of bosons. 
A particularly well characterized magnet is 
yttrium-iron garnet (YIG) \cite{Cherepanov93}.
The momentum distribution function of the 
magnon gas in YIG has been probed experimentally using the technique of 
Brillouin light  scattering \cite{Demokritov06,Demidov08a,Demidov08b,Sandweg10}.

At long wavelengths the energy dispersion of the experimentally relevant magnon band 
in thin YIG-stripes can be approximated by Eq.~(\ref{eq:ekyig}).
The values of the exchange stiffness $\rho_{\rm ex}$ and the dipolar energy scale
$\Delta$ entering  Eq.~(\ref{eq:ekyig})  can be written as
$\rho_{\rm ex} = JS a^2$ and $\Delta = 4 \pi \mu M_s$, where
$J \approx 1.29\, \mathrm{K}$ is the
nearest neighbor exchange constant of YIG and the saturation
magnetization is given by $ 4 \pi M_s \approx 1750\,  {\rm G}$.
Here $a \approx 12.376\, \mathrm{\AA}$ is the lattice constant and 
$\mu = g \mu_B $ is the magnetic moment
of the localized spins \cite{Kreisel09}.
If we arbitrarily set the effective $g$-factor equal to two so that
$\mu = 2 \mu_B$, then the effective spin is
$S = M_s a^3/ \mu \approx 14.2$.
Due to an interplay between exchange interactions, dipole-dipole interactions, and
finite-size effects, the magnon dispersion of YIG exhibits two degenerate minima
at finite momenta $\pm k_{\rm min} \hat{\bd{z}}$, where
 $\hat{\bd{z}}$ is the direction of the external magnetic field.
Having fixed the energy dispersion, 
the only adjustable parameters of our model Hamiltonian (\ref{eq:hdef})
are the phonon velocity $c = \omega_{\bd{q}} / | \bd{q} |$
and the magnon-phonon coupling constant $\gamma = \gamma_{\bd{q}} /
 \sqrt{ | \bd{q} | } $.

Highly excited nonequilibrium states of the magnon gas in YIG can be generated
via  parametric pumping with a time-dependent external microwave source.
Once the microwave source has been switched off, the thermalization
of the magnon gas can be directly observed\cite{Demidov08a,Demidov08b}.
In particular, in Ref.~[\onlinecite{Demidov08b}] the time evolution of the momentum 
distribution function $n_{\bd{k}} ( t )$ of the magnon gas in YIG 
has been measured. The data presented by Demidov {\it{et al.}} \cite{Demidov08b}
show how a highly excited initial state evolves towards a thermal equilibrium 
state where the occupation of 
the magnon modes at the bottom of the spectrum is strongly enhanced.
Whether or not such a state should be called a Bose-Einstein condensate of 
magnons seems to be a matter of semantics \cite{Snoke06,Bugrij07}.
We have argued previously\cite{Hick10,Rueckriegel12} that 
the experimentally observed strong enhancement of
the magnon distribution is not accompanied by superfluidity, because
a macroscopic occupation of a certain magnon mode  is simply equivalent
with a change in magnetic order \cite{Kohn70}.

In Fig.~\ref{fig:thermalization} we show
our results for the time evolution of the magnon distribution
obtained from the numerical solution
of the FRG rate equations ({\ref{eq:SigmaKFRG}, \ref{eq:SigmaIFRG}) 
and the particle number conserving kinetic equation
given in Eqs.~(\ref{eq:kinmastermu}, \ref{eq:mures}).
For the numerical calculations 
we have used the energy dispersion (\ref{eq:ekyig}) with experimentally relevant
parameters discussed above and stripe thickness 
$d = 5 \, \mathrm{\mu m }$. The minima of the dispersion are located at 
\mbox{$| k_{\parallel}  | = 5.5\times10^4\,\mathrm{cm}^{-1} 
\equiv k_{\rm{min}}$} and \mbox{$k_{\bot} = 0$}.
For simplicity we consider a square stripe 
with the length of $L = 4.58 \,\mathrm{\mu m}$ and work
with a truncated momentum space such that
\mbox{$|k_{\parallel}| \leq  41.1\times10^4\,\mathrm{cm}^{-1}$}
and \mbox{$|k_{\bot} | \leq  13.7\times10^4\,\mathrm{cm}^{-1}$}, 
corresponding to $61\times21$ points in momentum space.
The rate equation and the flow equations are solved simultaneously as described in 
Ref.~[\onlinecite{Kloss11}]. In order to calculate the imaginary part of the 
phonon self-energy as discussed in Secs.~\ref{sec:rate} and \ref{sec:FRG} we
have used a two-dimensional adaption of the tetrahedron method\cite{Jepsen78}.
The initial cutoff $\Lambda_0$ was chosen to be 
$404\, \Omega_0$, where 
\mbox{$ \Omega_0 = c k_{\mathrm{min}} $} is the relevant energy scale for the
phonon dynamics.
We have set the phonon velocity in YIG to 
\mbox{$c = 7.15 \times 10^5 \, \mathrm{cm/s}$} as measured in 
Ref.~[\onlinecite{Siu01}].
For the numerical calculations we have divided the time interval into $3240$ equidistant steps up to a final time of $670\,\mathrm{ns}$.
For small values of the flow parameter $\Lambda \lesssim \Omega_0$ 
the dependence of the magnon self-energies on $\Lambda$ is rather strong. 
We have therefore divided the range of $\Lambda$ into two intervals, $[0,4\Omega_0]$ and $[4\Omega_0, \Lambda_0]$, each equidistantly discretized with $500$ points.
%
%
\begin{figure}[tb]
  \centering
\includegraphics[width=0.45\textwidth]{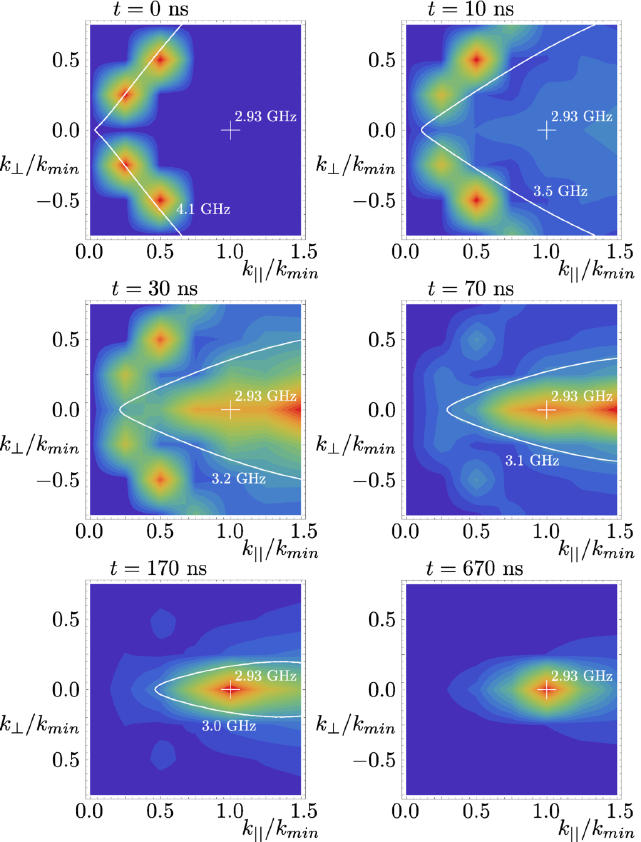}\nspace
\vspace{5mm}
  \vspace{-4mm}
  \caption{%
(Color online)
These contour plots show an interpolation of the magnon distribution of YIG on a 
$7\times7$-mesh in momentum space for different 
times. The white contour lines denote the magnon spectrum
of YIG for fixed  energies. 
The crosses mark the minimum of the  dispersion.
The results are presented such that they are directly  comparable with the experimental 
data of Ref.~[\onlinecite{Demidov08b}]. 
We have shifted the zero of time  to the moment where  
the microwave pumping is switched off.
For the calculation of the dispersion of YIG
we have used the parameters from Ref.~[\onlinecite{Kreisel09}],
which differ slightly from the parameters used in Ref.~[\onlinecite{Demidov08b}].
The initial state is chosen such that
magnons with an energy of $4.1\, \mathrm{GHz}$ in the
area $\left| k_{\bot} \right| < 0.5 k_{{\rm{min}}}$ are equally occupied, which
resembles the initial magnon distribution in the 
experiment after the pumping has been switched off.
The strength of the  magnon-phonon interaction has been adjusted such 
that the occupation of the ground state is saturated after approximately 
$670 \, \mathrm{ns}$, which gives 
$\gamma = 2.0 \times 10^{-7} \, \mathrm{cm\sqrt{GHz}} $. 
Finally, we have used typical experimental values
for temperature  ($T = 300 \, \mathrm{K}$), magnetic field ($H = 1000 \, \mathrm{Oe}$),
and magnon  density ($4.77 \times 10^6 / \mathrm{\mu m}^2$).
}
\label{fig:thermalization}
\end{figure}
Our theoretical results shown in Fig.~\ref{fig:thermalization}
agree quite well with the corresponding experimental
data shown in Fig.~3 of Ref.~[\onlinecite{Demidov08b}].
Note that in our simple model the order parameter 
$  \langle a^C_{\bd{k}} ( t ) \rangle$ vanishes, so that
the global $U(1)$ symmetry of the Hamiltonian (\ref{eq:hdef}) is not
spontaneously broken and our magnon gas is not superfluid.
Nevertheless, our model describes
the experimentally  observed thermalization towards
a magnon state with strongly enhanced occupation
at the bottom of the spectrum, indicating that the experiments observing BEC of 
magnons in YIG can be explained without postulating that
the magnon state is superfluid.

Of course,  the thermalization of the magnon gas in YIG 
is not exclusively driven by magnon-phonon interactions, but also by
many-body scattering processes involving two or more magnons.
In particular, interactions involving 
three and four powers of the magnon operators should be added
to obtain a more realistic model for the magnon gas in YIG\cite{Hick10}.
While the three-magnon vertices 
yield the dominant contribution to the magnon damping \cite{Chernyshev12},
the four-magnon vertices 
compete with the magnon-phonon interaction
to re-distribute the excited initial state over 
the  magnon spectrum.
The inclusion of these scattering processes into our FRG formalism is
in principle possible but requires
substantial extensions of the present formalism which are
beyond the scope of this work. 
On the other hand, experiments show that magnons in YIG behave as well-defined, weakly interacting quasiparticles, even in an external microwave field. Our assumption
that the initial state at $t = 0$ is not correlated is therefore well justified 
for YIG.

Finally, let us point out that 
the magnon-phonon interaction in our
model Hamiltonian (\ref{eq:hdef})
conserves the number of magnons
because the  phonon field $X_{\bd{q}}$
couples only to the magnon density $\hat{\rho}_{\bd{q}}$.
However, we expect that in YIG there should
exist another type of  magnon-phonon interaction involving the combinations
$ a^{\dagger}_{\bd{k}} a^{\dagger}_{ - \bd{k} - \bd{q}} X_{\bd{q}}   $
and 
$ a_{-\bd{k}} a_{  \bd{k} +  \bd{q}} X_{-\bd{q}} $.
This type of magnon-phonon coupling does not conserve the magnon number
and leads to an efficient energy transfer from the magnons  
into the phonon system.
Microscopically, this  magnon-phonon interaction arises from the fact that
the quasi-particle  operators $a_{\bd{k}}$ describing the
physical magnon excitations of YIG are related via a canonical (Bogoliubov)
transformation 
to the Holstein-Primakoff  bosons resulting from the bosonization
of the spin operators;
see Ref.~[\onlinecite{Kreisel11}] for a microscopic derivation of these terms
for a frustrated antiferromagnet.
Even if the couplings of these processes in YIG are small,
we expect that these processes are important to describe the experimentally 
observed\cite{Dzyapko11} long-time decay of the BEC.

\section{Conclusions}

In summary, we have studied the thermalization of the magnon gas 
due to coupling to the thermal phonon bath
in thin films of the magnetic insulator yttrium-iron garnet.
Although we have only retained the interaction processes 
which conserve the total number of
magnons, our theoretical predictions for the time evolution of the
momentum distribution of the magnons agrees quite well with 
experimental results. In particular, the accumulation of magnons at the bottom 
of the spectrum is correctly described within our approach.

We have intentionally considered in this work only the case 
where
all  nonequilibrium averages  $ \langle a_{\bd{k}}^C ( t ) \rangle $
vanish identically\cite{footnoteHaug}. These averages play the role of order parameters;
keeping in mind that the magnon operators in YIG describe the quantized fluctuations
around the classical ground states,
macroscopic values of
 $ \langle a_{\bd{k}}^C ( t ) \rangle $
for one or several wave-vectors describe a macroscopic  re-organization
of the magnetic state \cite{Rueckriegel12}. 
To describe this phenomenon microscopically, it is necessary to take
the magnon-magnon interactions into account.

The present work describes also some technical advance: we have developed 
an implementation of the nonequilibrium 
functional renormalization group method
which clearly exhibits the connection to a conventional rate equation approach.
In a sense, our FRG approach can be considered as
a simple  renormalization group extension of a rate equation approach.
The main advantages  of our approach are that  our FRG resummation
goes beyond the simple second order approximation for the
collision terms in the kinetic equation, and that our method  allows us to
include in a simple way the feedback of the
magnon dynamics on the phonon propagators which in turn determine the
transition rates between the magnon states in the rate equation.
We have also developed a generalized hybridization cutoff scheme 
for interacting bosons.
It is possible to generalize our approach to include magnon-magnon interactions 
or magnon number violating magnon-phonon interactions.

\section*{ACKNOWLEDGMENTS}
We thank Mathieu Taillefumier for his help with the
numerical calculations, and Sasha Chernyshev for sending us
his recent  preprint on magnon damping in YIG prior to publication.
We also thank  Andreas Kreisel and Oleksandr Serha for helpful discussions. 
Financial support by
SFB/TRR49,  the CNRS, and the Humboldt foundation is gratefully
acknowledged.

\begin{appendix}

\renewcommand{\theequation}{A\arabic{equation}}
\section*{APPENDIX A: Quantum kinetic equations}
\setcounter{equation}{0}

In this appendix we review several possibilities 
of writing down quantum kinetic equations.
Although in the main text we only make use of the Wigner transformed
kinetic equation for the distribution function $g_{\bd{k}} ( \tau ; \omega )$
defined via Eq.~(\ref{eq:Gstruc}), it is instructive
to work out the precise relation between our parametrization and 
a formulation of the nonequilibrium problem in terms of
the two-time Keldysh Green function.

\subsection{Two-time Keldysh Green function}

Writing the nonequilibrium Dyson equation (\ref{eq:dyson}) as
\begin{equation}
   \left( \mathbf{G}_0^{-1} -  \mathbf{\Sigma} \right)  \mathbf{G}  =  \mathbf{I} 
\end{equation}
we obtain following three
equations for the sub-blocks,
 \begin{subequations}
 \begin{eqnarray}
    ( \hat{G}_{0}^{R} )^{-1} \hat{G}^R & = & {\hat{I}}  + \hat{\Sigma}^R 
\hat{G}^R   ,
 \label{eq:DysonR}
 \end{eqnarray}
 \begin{eqnarray}
    ( \hat{G}_{0}^{A} )^{-1} \hat{G}^A & = & {\hat{I}} +  \hat{\Sigma}^A  \hat{G}^A    ,
 \label{eq:DysonA}
 \end{eqnarray}
 \begin{eqnarray}
 ( \hat{G}_{0}^{R} )^{-1} \hat{G}^K
 & = &    \hat{\Sigma}^K \hat{G}^A  + \hat{\Sigma}^R  \hat{G}^K   ,
 \label{eq:GKdyson}
 \end{eqnarray}
 \end{subequations}
where $\hat{I}$ is the unit matrix in
the momentum- and  time labels.
Alternatively we can also consider the corresponding
``right Dyson equation'',
\begin{equation}
   \mathbf{G} \left( \mathbf{G}_0^{-1} -  \mathbf{\Sigma} \right)  =  \mathbf{I} ,
\end{equation}
which implies the following relations,
 \begin{subequations}
 \begin{eqnarray}
  \hat{G}^R     ( \hat{G}_{0}^{R} )^{-1}  & = & {\hat{I}}
 + \hat{G}^R    \hat{\Sigma}^R ,
 \label{eq:DysonRright}
 \end{eqnarray}
 \begin{eqnarray}
  \hat{G}^A   ( \hat{G}_{0}^{A} )^{-1}  & = & {\hat{I}} 
 + \hat{G}^A   \hat{\Sigma}^A ,
 \label{eq:DysonAright}
 \end{eqnarray}
 \begin{eqnarray}
   \hat{G}^K  ( \hat{G}_{0}^{A} )^{-1}
 & = &
    \hat{G}^R \hat{\Sigma}^K
 + \hat{G}^K  \hat{\Sigma}^A .
 \label{eq:GKdysonright}
 \end{eqnarray}
 \end{subequations}
Subtracting the Keldysh component of the left and right-hand sides of the Dyson equations  (\ref{eq:GKdyson},\ref{eq:GKdysonright}),  we obtain the kinetic equation
 \begin{equation}
 \bigl[ \hat{M}_0 , \hat{G}^K \bigr] = 
 \hat{\Sigma}^R \hat{G}^K
  - \hat{G}^R \hat{\Sigma}^K  + \hat{\Sigma}^K \hat{G}^A
  - \hat{G}^K \hat{\Sigma}^A,
 \label{eq:kinG}
 \end{equation}
where $ [\;  , \; ] $ denotes the commutator, and
$\hat{M}_0$ is defined in Eq.~(\ref{eq:M0def}). Note that
\begin{equation}
  ( {\hat{G}_0^{R}} )^{-1}  =  \hat{M}_0 + i \eta \hat{I}
 \quad , \quad
  ( {\hat{G}_0^{A}} )^{-1}  =   \hat{M}_0 - i \eta \hat{I} ,
 \label{eq:GRAM}
 \end{equation}
and
 \begin{subequations}
 \begin{eqnarray}
  {[} \hat{M}_0 \hat{G}^K ]_{ t t^{\prime}} & = &  ( i \partial_t - \epsilon_{\bd{k}} ) 
 {G}^K ( t,  t^{\prime}),
 \\
 {[} \hat{G}^K \hat{M}_0   ]_{ t t^{\prime}} & = &  ( - i \partial_{t^{\prime}} 
- \epsilon_{\bd{k}} )  {G}^K ({t, t^{\prime}}),
 \end{eqnarray}
 \end{subequations}
implying
 \begin{eqnarray} 
 {[} \hat{M}_0 ,  \hat{G}^K       ]_{ t t^{\prime} } 
   =    ( i \partial_t + i \partial_{ t^{\prime} } ) G^K ({ t, t^{\prime}}).
 \end{eqnarray}
We conclude that in the time domain the quantum kinetic equation  (\ref{eq:kinG})
assumes the form 
\begin{eqnarray}
 & &  ( i \partial_t + i \partial_{t^{\prime}} ) G^{K} ( t , t^{\prime} ) 
 \nonumber
 \\
& = & \int_{t_0}^t d t_1 [ \Sigma^R ( t, t_1 ) G^K ( t_1 , t^{\prime} )
 -  G^R ( t, t_1 ) \Sigma^K ( t_1 , t^{\prime} ) ]
 \nonumber
 \\
&+ & 
\int_{t_0}^{t^{\prime}} d t_1 [ \Sigma^K ( t, t_1 ) G^A ( t_1 , t^{\prime} )
 -  G^K ( t, t_1 ) \Sigma^A ( t_1 , t^{\prime} ) ].
 \nonumber
 \\
 & &
 \end{eqnarray}
To identify the operators which reduce in the appropriate limit to
the collision integral in the Boltzmann equation,
it is convenient to introduce\cite{Rammer07} 
$\hat{\Sigma}^M  =  \frac{1}{2}[ \hat{\Sigma}^R +  \hat{\Sigma}^A ]$ and
$ \hat{\Sigma}^I  =  i [ \hat{\Sigma}^R -  \hat{\Sigma}^A ]$, see Eq.~(\ref{eq:Asymdef}).
The inverse relations are
\begin{equation}
 \hat{\Sigma}^R  =   \hat{\Sigma}^M - \frac{i}{2}  \hat{\Sigma}^I
\quad , \quad
\hat{\Sigma}^A  =   \hat{\Sigma}^M + \frac{i}{2}  \hat{\Sigma}^I .
 \label{eq:SigmaAMI}
 \end{equation}
A similar decomposition is also introduced for the
retarded and advanced Green functions,
 \begin{equation}
\hat{G}^M  =  \frac{1}{2} [  \hat{G}^R + \hat{G}^A  ]
\quad , \quad
 \hat{G}^I  =  i [ \hat{G}^R - \hat{G}^A ],
 \label{eq:GIdef}
 \end{equation}
so that
 \begin{equation}
 \hat{G}^R  =  \hat{G}^M - \frac{i}{2} \hat{G}^I
 \quad , \quad
\hat{G}^A  =  \hat{G}^M + \frac{i}{2} \hat{G}^I .
 \end{equation}
Defining  $\hat{M} = \hat{M}_0 - \hat{\Sigma}^M$
[see Eq.~(\ref{eq:Mdef})], we may 
write the kinetic equation (\ref{eq:kinG}) as
 \begin{equation}
 \bigl[ \hat{M} , \hat{G}^K \bigr] -
\bigl[  \hat{\Sigma}^K , \hat{G}^M \bigr] = \hat{C}^{\rm in} - 
\hat{C}^{\rm out},
 \label{eq:kineqGK}
 \end{equation}
where we have introduced the collision matrices
 \begin{eqnarray}
 \hat{C}^{\rm in} & = & \frac{i}{2} \bigl\{    \hat{\Sigma}^K ,  \hat{G}^I
\bigr\}, 
\\
\hat{C}^{\rm out} & = & \frac{i}{2} \bigl\{ \hat{\Sigma}^I , \hat{G}^K 
\bigr\}.
 \end{eqnarray}
Recall that  $\{\; , \; \}$ denotes the  anticommutator.
The notation indicates that after the usual approximations 
the matrix elements of 
of $\hat{C}^{\rm in}$ and $\hat{C}^{\rm out}$
can be identified with the in-scattering and the out-scattering contributions to the collision
integral  in the Boltzmann equation which emerges from Eq.~(\ref{eq:kineqGK})
in the classical limit.

\subsection{Distribution function}

Another form of the kinetic equation can be obtained by 
expressing $\hat{G}^K$ in terms of the distribution function $\hat{g}$ by setting
\begin{equation}
 \hat{G}^K = \hat{G}^R \hat{g}^{\dagger} - \hat{g} \hat{G}^A,
 \label{eq:Gstruc2}
 \end{equation}
see Eq.~(\ref{eq:Gstruc}).
Note that a similar ansatz by Kamenev\cite{Kamenev04,Kamenev11}
assumes that $\hat{g}$ is hermitian, which in general is not correct.
A formal proof that $\hat{G}^K$ indeed can be written in the form (\ref{eq:Gstruc})
can be found in Ref.~[\onlinecite{Lipavsky86}]
(see also  Appendix~B of Ref.~[\onlinecite{Kloss11}]).
The ansatz (\ref{eq:Gstruc2}) has some formal similarity to the
generalized Kadanoff-Baym ansatz \cite{Lipavsky86}
 \begin{equation}
 G^{K} ( t , t^{\prime} ) = i [ G^{R}  ( t , t^{\prime} ) G^{K} ( t^{\prime} , t^{\prime} )
 - G^{K} ( t , t ) G^A ( t , t^{\prime} ) ]    ,
 \label{eq:GKB}
 \end{equation}
which is often used to express the Keldysh Green function at different times
in terms of its equal-time matrix elements.
In fact, the generalized Kadanoff-Baym ansatz amounts to setting
 \begin{equation}
 [ \hat{g} ]_{ t t^{\prime}} =   [ \hat{g}^{\dagger} ]_{ t t^{\prime}}  = \delta ( t - t^{\prime} ) i G^K ( t , t ).
 \end{equation} 
Substituting Eq.~(\ref{eq:Gstruc2}) into our
definition (\ref{eq:SigmaKdef}) of the Keldysh self-energy $\hat{\Sigma}^K$,
we obtain the kinetic equation
 \begin{equation}
  \hat{M}_0 \hat{g} - \hat{g}^{\dagger} \hat{M}_0 = - \hat{\Sigma}^{K} + 
 \hat{\Sigma}^{R} \hat{g} - \hat{g}^{\dagger} \hat{\Sigma}^A ,
 \label{eq:kineqgunsub}
 \end{equation}
which can also be written as [see Eq.~(\ref{eq:kineqg})]
 \begin{equation}
  - i ( \hat{M} \hat{g} - \hat{g}^{\dagger} \hat{M} ) = \hat{\Sigma}^{\rm in} - 
 \hat{\Sigma}^{\rm out},
 \label{eq:kineqg2}
 \end{equation}
where  $\hat{\Sigma}^{\rm in}$ and
$ \hat{\Sigma}^{\rm out}$ are defined  in Eqs.~(\ref{eq:Sigmain}, \ref{eq:Sigmaout}).
Note that in the non-interacting limit
or in a simple approximation where interaction corrections to the Keldysh self-energy
$\hat{\Sigma}^K$ are neglected, we see from
Eqs.~(\ref{eq:regg}) and (\ref{eq:SigmaKdef}) that
$\hat{\Sigma}^K$ reduces to an infinitesimal regularization
$-  i \eta ( \hat{g} + \hat{g}^{\dagger}) $.

\subsection{Wigner transformed Keldysh Green function}

To describe the approach to equilibrium, it can be advantageous
to perform a partial Fourier transformation of the two-time Green functions
with respect to the time difference (Wigner transformation), see Eq.~(\ref{eq:Wignerdef}).
Using the fact that
 \begin{eqnarray} 
  & & {[} \hat{M}_0 \hat{G}^K - \hat{G}^K \hat{M}_0       ]_{ t t^{\prime} } 
   =    ( i \partial_t + i \partial_{ t^{\prime} } ) G^K ({ t, t^{\prime}})
 \nonumber
 \\
 &     & = 
 \int_{ - \infty}^{\infty}   \frac{ d \omega}{2 \pi} 
 e^{  - i \omega  ( t - t^{\prime})  }  i \partial_\tau G^K ( \tau ; \omega ) ,
 \end{eqnarray}
where  ${G}^K ( \tau ; \omega )$ is the Wigner transform of 
${G}^K( t , t^{\prime})$, we obtain the following expression for
the Wigner transform of our kinetic equation (\ref{eq:kineqGK})
for the Keldysh Green function,
 \begin{eqnarray}
& &  i \partial_\tau G^K ( \tau; \omega ) -  [ \hat{\Sigma}^M, \hat{G}^K  ]_{ ( \tau ; \omega )}
-   [  \hat{\Sigma}^K ,   \hat{G}^M ]_{ ( \tau ; \omega )}
 \nonumber
 \\
& &
  =
 C^{\rm in} ( \tau ; \omega ) - C^{\rm out} ( \tau ; \omega ),
\label{eq:kinGT}
 \end{eqnarray}
where
 \begin{eqnarray}
  C^{\rm in} ( \tau ; \omega ) & = & 
\frac{i}{2} \bigl\{    \hat{\Sigma}^K ,   \hat{G}^I
\bigr\}_{ ( \tau ; \omega )}, 
\\
{C}^{\rm out} ( \tau ; \omega )    & = & \frac{i}{2} \bigl\{   \hat{\Sigma}^I ,   \hat{G}^K
\bigr\}_{ ( \tau ; \omega )}.
 \end{eqnarray}

Recall that the Wigner transform of the product of two matrices $\hat{A}$ and $\hat{B}$ 
in the time labels can be expressed in terms of the
Wigner transforms of the factors $\hat{A}$ and $\hat{B}$ as\cite{Kamenev11}
 \begin{eqnarray}
 {[} \hat{A} \hat{B} {]}_{( \tau ; \omega )} & = & \int d  s_1 \int d s_2 
 \int  \frac{ d \omega_1}{ 2 \pi} \int \frac{ d \omega_2}{ 2 \pi}
e^{ i ( \omega_1 s_2 - \omega_2 s_1 )}
 \nonumber
 \\
 &  \times &
 A( \tau + \frac{s_1}{2} ;  \omega + \omega_1  )
 B (  \tau + \frac{s_2}{2} ;   \omega + \omega_2 ).\hspace{7mm}
 \end{eqnarray}
Expanding the arguments of  $A$ and $B$ for small $\omega_1$, $\omega_2$, 
$s_1$, $s_2$,  and retaining only the first order
in the derivatives we obtain the approximate expression
 \begin{eqnarray}
 {[} \hat{A} \hat{B}{]}_{( \tau; \omega )} & \approx & A ( \tau ; \omega ) B ( \tau; \omega )
 \nonumber
 \\
 &  & \hspace{-20mm} + \frac{1}{2i} \left[
 \partial_\tau A ( \tau ; \omega )
\partial_{\omega} B ( \tau ; \omega )
 - 
 \partial_{\omega} A ( \tau ; \omega )
 \partial_\tau B ( \tau ; \omega ) \right]. \hspace{7mm}
 \label{eq:WTprod}
 \end{eqnarray}
Assuming that the $\tau$-dependence of the distribution functions
is slow, we may
retain only the first term  in the right-hand side of Eq.~(\ref{eq:WTprod}), i.e. we
approximate the Wigner transforms of the products appearing 
in the kinetic equations by the products of the Wigner transforms of the factors.
In this approximation the Wigner transforms of the
commutator terms
on the left-hand side of the kinetic equation (\ref{eq:kinGT}) for
 $G^K ( \tau; \omega )$ vanish, while the
Wigner transforms of the collision integrals are
 \begin{eqnarray}
 C^{\rm in} ( \tau ; \omega ) & = & i \Sigma^K ( \tau; \omega ) G^I ( \tau ; \omega ),
 \\
C^{\rm out} ( \tau ; \omega ) & = & i \Sigma^I ( \tau; \omega ) G^K ( \tau ; \omega ).
 \end{eqnarray}
Given $G^{K} ( \tau ; \omega )$, we still have to perform a frequency
integration to reconstruct the equal time Keldysh Green function (i.e.
the distribution function),
 \begin{equation}
 G^K ( t , t ) = \int_{- \infty}^{\infty} \frac{d \omega}{2 \pi} G^K ( t ; \omega ).
 \end{equation}

\subsection{Wigner transformed distribution function}

We can avoid the above frequency integration by working
with the Wigner transform $g_{\bd{k}} ( \tau ; \omega )$
of the kinetic equation (\ref{eq:kineqg})
for the distribution matrix $\hat{g}$,
which is given in Eq.~(\ref{eq:kingWig}).
 Note that in thermal equilibrium and
in the absence of interactions the function $g_{\bd{k}} ( \tau ; \omega )$
is simply given by a symmetrized Bose function
 \begin{equation}
 g_{\bd{k}} ( \tau ; \omega ) =  1 + \frac{2}{e^{\beta ( \omega - \mu )} -1 } 
= \coth \left[ \frac{\beta}{2} ( \omega - \mu ) \right]  .
 \end{equation}
In an interacting system with well-defined quasi-particles, we expect
that $g_{\bd{k}} ( \tau ; \omega )$ will relax towards
a $\tau$-independent function $g_{\bd{k}} ( \omega )$ which for
$\omega $ close to the renormalized quasi-particle energy can be expressed in terms of a symmetrized Bose function
with some effective temperature. 
Using
 \begin{eqnarray}
 {[}   \hat{M}_0 \hat{g} - \hat{g}^{\dagger} \hat{M}_0       ]_{ t t^{\prime} }   & = & 
 i \partial_{ t} g_{ t t^{\prime}} +  i  \partial_{t^{\prime}} g^{\ast}_{ t^{\prime} t } -
 \epsilon_{\bd{k}} ( g_{ t t^{\prime}} - g^{\ast}_{ t^{\prime} t } )
 \nonumber
 \\
 &    = &
 \int_{ - \infty}^{\infty}   \frac{ d \omega}{2 \pi} 
  e^{  - i \omega  ( t - t^{\prime})  } 
 \Bigl[  i \partial_\tau {\rm Re} g  ( \tau ; \omega )  
\nonumber
 \\
 & &
 + 2 i ( \omega - \epsilon_{\bd{k}} ) {\rm Im} g ( \tau ; \omega )
 \Bigr],
 \label{eq:drivingWignerg}
 \end{eqnarray}
we finally arrive at Eq.~(\ref{eq:kingWig}).

\renewcommand{\theequation}{B\arabic{equation}}
\section*{APPENDIX B: Second order self-energies}
\setcounter{equation}{0}

For the derivation of the perturbative kinetic equation
(\ref{eq:master}) we need the components of the nonequilibrium self-energy
to second order in the magnon-phonon coupling.
In the contour-basis the self-energies are
 \begin{eqnarray}
 \Sigma^{ p p^{\prime}}_{\bd{k}} ( t , t^{\prime} ) & = & p p^{\prime} 
  \frac{i}{V}\sum_{\bd{q}}  \gamma_{\bd{q}}^2
 \Bigl[ D^{p p^{\prime}}_{\bd{q}} ( t , t^{\prime} )
 G^{p p^{\prime}}_{\bd{k} - \bd{q}} ( t , t^{\prime} )
 \nonumber
 \\
 & &
\hspace{20mm} + ( \bd{q} \rightarrow - \bd{q} ) \Bigr].
 \end{eqnarray}
Here the $p, p^{\prime} \in \{ +, - \}$ label the two branches of the
Keldysh contour.
Using the fact that the $( \bd{q} \rightarrow - \bd{q})$ term gives just a factor of
$2$ because the Green functions are even functions of the momentum argument,
we obtain in the Keldysh basis
 \begin{subequations}
 \begin{eqnarray}
 \Sigma^{ R}_{\bd{k}} ( t , t^{\prime} ) & = & 
\Sigma^{ QC}_{\bd{k}} ( t , t^{\prime} ) =
  \frac{i}{V}\sum_{\bd{q}}  \gamma_{\bd{q}}^2
 \Bigl[ D^{R}_{\bd{q}} ( t , t^{\prime} )
 G^{K}_{\bd{k} - \bd{q}} ( t , t^{\prime} )
\nonumber
 \\
 & &
 \hspace{20mm}
+ D^{K}_{\bd{q}} ( t , t^{\prime} )
 G^{R}_{\bd{k} - \bd{q}} ( t , t^{\prime} ) \Bigr],
 \label{eq:SigmaRtt}
 \\
 \Sigma^{ A}_{\bd{k}} ( t , t^{\prime} ) & = & 
\Sigma^{ CQ}_{\bd{k}} ( t , t^{\prime} ) =
  \frac{i}{V}\sum_{\bd{q}}  \gamma_{\bd{q}}^2
 \Bigl[ D^{A}_{\bd{q}} ( t , t^{\prime} )
 G^{K}_{\bd{k} - \bd{q}} ( t , t^{\prime} )
\nonumber
 \\
 & & \hspace{20mm}
+ D^{K}_{\bd{q}} ( t , t^{\prime} )
 G^{A}_{\bd{k} - \bd{q}} ( t , t^{\prime} ) \Bigr],
 \\
 \Sigma^{ K}_{\bd{k}} ( t , t^{\prime} ) & = & 
\Sigma^{ QQ}_{\bd{k}} ( t , t^{\prime} ) =
  \frac{i}{V}\sum_{\bd{q}}  \gamma_{\bd{q}}^2
 \Bigl[ D^{K}_{\bd{q}} ( t , t^{\prime} )
 G^{K}_{\bd{k} - \bd{q}} ( t , t^{\prime} )
\nonumber
 \\
 & & \hspace{-13mm}
+ D^{R}_{\bd{q}} ( t , t^{\prime} )
 G^{R}_{\bd{k} - \bd{q}} ( t , t^{\prime} )
+
 D^{A}_{\bd{q}} ( t , t^{\prime} )
 G^{A}_{\bd{k} - \bd{q}} ( t , t^{\prime} )
  \Bigr].
 \end{eqnarray}
\end{subequations}
The corresponding diagrams are shown in Fig.~\ref{fig:diagramsself}.
\begin{figure}[tb]
  \centering
\includegraphics[width=0.4\textwidth]{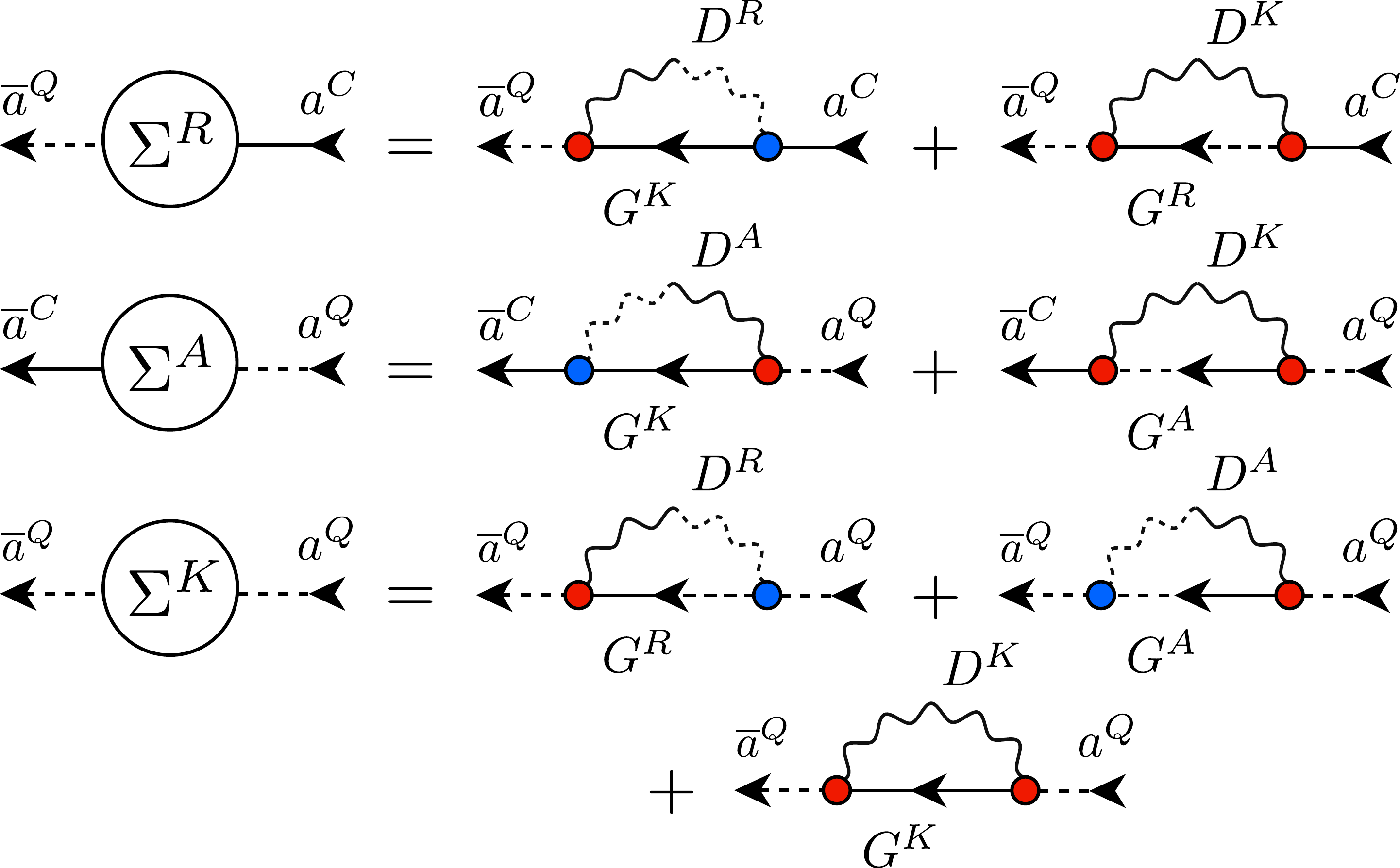}\nspace
\vspace{5mm}
  \vspace{-4mm}
  \caption{%
(Color online)
Diagrams contributing to the magnon self-energies to second order
in the magnon-phonon coupling. The symbols are defined in the caption
of Fig.~\ref{fig:flowself}. 
}
\label{fig:diagramsself}
\end{figure}
The spectral component of the self-energy  is
 \begin{eqnarray}
 \Sigma^{I}_{\bd{k}} ( t , t^{\prime} ) & = & i [ 
 \Sigma^{R}_{\bd{k}} ( t , t^{\prime} )
 - \Sigma^{A}_{\bd{k}} ( t , t^{\prime} ) ] 
 \nonumber
 \\
 &  = & \frac{i}{V}\sum_{\bd{q}}  \gamma_{\bd{q}}^2
 \Bigl[ D^{I}_{\bd{q}} ( t , t^{\prime} )
 G^{K}_{\bd{k} - \bd{q}} ( t , t^{\prime} )
\nonumber
 \\
 & & \hspace{13mm}
+ D^{K}_{\bd{q}} ( t , t^{\prime} )
 G^{I}_{\bd{k} - \bd{q}} ( t , t^{\prime} ) \Bigr].
 \end{eqnarray}
Moreover, using the fact that the
product of a retarded and an advanced function with the same time argument vanishes,
the Keldysh component of the self-energy can alternatively be written as
 \begin{eqnarray}
 \Sigma^{ K}_{\bd{k}} ( t , t^{\prime} ) & = & 
  \frac{i}{V}\sum_{\bd{q}}  \gamma_{\bd{q}}^2
 \Bigl[ D^{K}_{\bd{q}} ( t , t^{\prime} )
 G^{K}_{\bd{k} - \bd{q}} ( t , t^{\prime} )
\nonumber
 \\
 & &  \hspace{13mm}
-  D^{I}_{\bd{q}} ( t , t^{\prime} )
 G^{I}_{\bd{k} - \bd{q}} ( t , t^{\prime} ) \Bigr].
 \end{eqnarray}
To derive Wigner transformed kinetic equations,
we use the fact that the Wigner transform of a product of two functions with the same
time arguments can be written as a convolution in frequency space,
 \begin{eqnarray}
& &  [ A ( t , t^{\prime} ) B ( t , t^{\prime}) ]_{ ( \tau, \omega )}
 \nonumber
 \\
& & =
 \int_{ - \infty}^{\infty} d s e^{ i \omega s} A( \tau + \frac{ s}{2} , \tau - \frac{s}{2} )
B( \tau + \frac{ s}{2} , \tau - \frac{s}{2} )
 \nonumber
 \\
&  & = \int_{ - \infty}^{\infty} \frac{d \omega^{\prime}}{2 \pi} A ( \tau ; \omega^{\prime} )
 B ( \tau ; \omega - \omega^{\prime} ).
 \end{eqnarray}
The Wigner transform of the spectral and Keldysh components of our self-energies
are therefore given by
 \begin{eqnarray}
 \Sigma^I_{\bd{k}} ( \tau ; \omega ) & = & 
 \frac{i}{V}\sum_{\bd{q}}   \int_{ - \infty}^{\infty} \frac{ d \omega^{\prime}}{2 \pi}
 \gamma_{\bd{q}}^2
 \Bigl[ D^{I}_{\bd{q}} ( \tau ; \omega^{\prime} )
 G^{K}_{\bd{k} - \bd{q}} ( \tau ; \omega - \omega^{\prime} )
\nonumber
 \\
 & &  \hspace{13mm}
+  D^{K}_{\bd{q}} ( \tau  ; \omega^{\prime} )
 G^{I}_{\bd{k} - \bd{q}} ( \tau ; \omega -  \omega^{\prime} ) \Bigr],
 \label{eq:SigmaI2}
 \\
 \Sigma^K_{\bd{k}} ( \tau ; \omega ) & = & 
 \frac{i}{V}\sum_{\bd{q}}   \int_{ - \infty}^{\infty} \frac{ d \omega^{\prime}}{2 \pi}
 \gamma_{\bd{q}}^2
 \Bigl[ D^{K}_{\bd{q}} ( \tau ; \omega^{\prime} )
 G^{K}_{\bd{k} - \bd{q}} ( \tau ; \omega - \omega^{\prime} )
\nonumber
 \\
 & &  \hspace{13mm}
-  D^{I}_{\bd{q}} ( \tau  ; \omega^{\prime} )
 G^{I}_{\bd{k} - \bd{q}} ( \tau ; \omega -  \omega^{\prime} ) \Bigr].
 \label{eq:SigmaK2}
 \end{eqnarray}

\renewcommand{\theequation}{C\arabic{equation}}
\section*{APPENDIX C: Derivation of the rate equation}
\setcounter{equation}{0}

In this appendix we summarize the approximations which are necessary 
to arrive  at the perturbative kinetic equation (\ref{eq:master}).
Using Eq.~(\ref{eq:Gstruc}) 
we may express the Keldysh component of the Green function
appearing on the right-hand side of Eqs.~(\ref{eq:SigmaI2}, \ref{eq:SigmaK2})
in terms of the Wigner transform  $g_{\bd{k}} 
( \tau ; \omega )$   of the distribution function.
To leading order in the gradients 
we may factorize the Wigner transform
of the matrix products in
Eq.~(\ref{eq:Gstruc}),
 \begin{equation}
 i G_{\bd{k}}^K ( \tau ; \omega ) \approx  G^{I}_{\bd{k}} ( \tau ; \omega ) g_{\bd{k}} 
( \tau ; \omega ).
 \label{eq:GK2}
 \end{equation}
After substituting the self-energies (\ref{eq:SigmaI2}) and (\ref{eq:SigmaK2})
with the Keldysh Green function given by Eq.~(\ref{eq:GK2}) into
the Wigner transformed kinetic equation (\ref{eq:kinqpfinal}),
we obtain an integro-differential
equation for the nonequilibrium distribution function
$g_{\bd{k}} ( \tau ) = g_{\bd{k}} ( \tau ; \omega = \epsilon_{\bd{k}} )$.
To further simplify this equation, let us assume that the
phonon system is in thermal equilibrium. Moreover, we approximate the
phonon propagators by the non-interacting equilibrium propagators.
The Wigner transforms of the phonon Green functions can then be replaced
by the usual Fourier transforms,
 \begin{subequations}
 \begin{eqnarray}
 F^R_{\bd{q}} ( \omega ) & = & \frac{1}{\omega - \omega_{\bd{q}} + i \eta },
 \\
 F^A_{\bd{q}} ( \omega ) & = & \frac{1}{\omega - \omega_{\bd{q}} - i \eta },
 \\
 F^I_{\bd{q}} ( \omega ) & = & i [ F^R_{\bd{q}} ( \omega ) -  F^A_{\bd{q}} 
 ( \omega ) ] = 2 \pi \delta ( \omega - \omega_{\bd{q}} ),
 \\
 i F^K_{\bd{q}} ( \omega ) & = & \left[ 1 + \frac{2}{e^{ \beta \omega} -1 }
 \right]  F^I_{\bd{q}} ( \omega )
  =  f^{0} ( \omega )  F^I_{\bd{q}} ( \omega ),
 \hspace{7mm}
 \end{eqnarray}
\end{subequations}
where 
 \begin{equation}
 f^0 ( \omega ) =  1 + \frac{2}{e^{ \beta \omega } -1 }
 = \coth \left( \frac{\beta \omega }{ 2} \right)
 \end{equation}
is the equilibrium phonon distribution.
The corresponding symmetrized phonon propagators are
\begin{subequations}
 \begin{eqnarray}
 D^R_{\bd{q}} ( \omega ) & = & 
\frac{1}{2}
 \left[  F^R_{\bd{q}} ( \omega ) +  F^A_{- \bd{q}} (- \omega ) \right]
 \nonumber
 \\
 & = & \frac{ \omega_{\bd{q}}}{ ( \omega + i \eta )^2 - \omega_{\bd{q}}^2 },
 \\
 D^A_{\bd{q}} ( \omega ) & = & \frac{1}{2}
 \left[  F^A_{\bd{q}} ( \omega ) +  F^R_{- \bd{q}} (- \omega ) \right]
 \nonumber
 \\
 & = & \frac{ \omega_{\bd{q}}}{ ( \omega - i \eta )^2 - \omega_{\bd{q}}^2 },
 \\
 D^I_{\bd{q}} ( \omega ) & = & 
 \frac{1}{2}
 \left[  F^I_{\bd{q}} ( \omega ) -  F^I_{- \bd{q}} (- \omega ) \right]
 \nonumber
 \\
 & = & \pi  {\rm sgn} (\omega )  [ \delta ( \omega - \omega_{\bd{q}} )
 + \delta ( \omega + \omega_{\bd{q}} ) ],
 \label{eq:DIpho}
 \\
 i D^K_{\bd{q}} ( \omega ) & = & 
  \frac{i}{2}
 \left[   F^K_{\bd{q}} ( \omega ) +   F^K_{- \bd{q}} (- \omega ) \right]
 \\
& = &  f^0 ( \omega  )    D^I_{\bd{q}} ( \omega )   .
 \label{eq:DKpho}
 \hspace{9mm}
 \end{eqnarray}
\end{subequations}
Substituting Eqs.~(\ref{eq:GK2}) and (\ref{eq:DKpho})
into Eqs.~(\ref{eq:SigmaI2}) and (\ref{eq:SigmaK2})
we obtain for the in-scattering and out-scattering components of the
nonequilibrium self-energy
 \begin{eqnarray}
   \Sigma^{\rm in}_{\bd{k}} ( \tau ; \omega ) &  = &  
 i \Sigma^{K}_{\bd{k}} ( \tau; \omega ) 
 \nonumber
 \\
 & = & 
 \frac{1}{ V} \sum_{\bd{q}} \gamma_{\bd{q}}^2 
 \int_{ - \infty}^{\infty} \frac{ d \omega^{\prime}}{2 \pi} D^I_{\bd{q}} ( \omega^{\prime} )
 G^I_{ \bd{k} - \bd{q}} ( \tau ; \omega - \omega^{\prime} )
 \nonumber
 \\
 & &  \times  \left[
 f^0 ( \omega^{\prime} ) 
 g_{\bd{k} - \bd{q}} ( \tau; \omega - \omega^{\prime} ) 
 + 1 \right] ,
 \\
 \Sigma^{\rm out}_{\bd{k}} ( \tau; \omega ) &  = & 
\Sigma^{I}_{\bd{k}} ( \tau; \omega )  
g_{\bd{k}}  ( \tau ; \omega )
 \nonumber
 \\
 & = &
 \frac{1}{ V} \sum_{\bd{q}} \gamma_{\bd{q}}^2 
 \int_{ - \infty}^{\infty} \frac{ d \omega^{\prime}}{2 \pi} D^I_{\bd{q}} ( \omega^{\prime} )
 G^I_{ \bd{k} - \bd{q}} ( \tau ; \omega - \omega^{\prime} )
 \nonumber
 \\
 & &  \times  \left[
 f^0 ( \omega^{\prime} ) +
 g_{\bd{k} - \bd{q}} ( \tau; \omega - \omega^{\prime} ) 
 \right]g_{\bd{k}}  ( \tau ; \omega ) .
 \hspace{7mm}
 \end{eqnarray}
Combining in- and out scattering contributions to the
collision integral, we obtain the quantum kinetic equation
\begin{eqnarray}
  \partial_{\tau} g_{\bd{k}} ( \tau ; \omega )  
 & = & \frac{1}{ V} \sum_{\bd{q}} \gamma_{\bd{q}}^2 
 \int_{ - \infty}^{\infty} \frac{ d \omega^{\prime}}{2 \pi} D^I_{\bd{q}} ( \omega^{\prime} )
 G^I_{ \bd{k} - \bd{q}} ( \tau ; \omega - \omega^{\prime} )
 \nonumber
 \\
 & \times & \Bigl\{
 f^0 ( \omega^{\prime} )  \left[
 g_{\bd{k} - \bd{q}} ( \tau; \omega - \omega^{\prime} ) - g_{\bd{k}} ( \tau ; \omega)   
 \right]
 \nonumber
 \\
 & &
 + 1 -  g_{\bd{k} - \bd{q}} ( \tau ; \omega - \omega^{\prime} ) g_{\bd{k}} ( \tau ;  \omega) 
 \Bigr\}.
 \label{eq:kinomega}
 \end{eqnarray}
In thermal equilibrium the different components of the magnon Green function
satisfy the fluctuation-dissipation theorem, which is equivalent with
 \begin{equation}
  g_{\bd{k}} ( \tau ; \omega ) \rightarrow \coth \left( \frac{ 
 \beta ( \omega - \mu )}{2}
 \right) = g^0 ( \omega ).
 \end{equation} 
Using the identity
 \begin{equation}
 \coth ( x-y ) = \frac{ 1 - \coth x \coth y}{ \coth x - \coth y },
 \end{equation}
with $ x = \beta ( \omega - \mu )/2$ and $y = \beta \omega^{\prime} /2$
it is easy to see that in thermal equilibrium the in- and out-scattering
contributions cancel, so that the  right-hand side 
of Eq.~(\ref{eq:kinomega}) vanishes identically.

To reduce Eq.~(\ref{eq:kinomega})  to the rate equation (\ref{eq:master})
we set
 \begin{eqnarray}
 g_{\bd{k}} ( \tau ; \omega ) & =  & 1 + 2 n_{\bd{k}} ( \tau ; \omega ),
 \\
 f^0 ( \omega ) & = & 1 + 2 b ( \omega ) = 1 + \frac{2}{ e^{\beta \omega} -1 },
 \end{eqnarray}
and shift the phonon momentum,  $\bd{q} = \bd{k} - \bd{k}^{\prime}$.
Then  Eq.~(\ref{eq:kinomega}) can be written as
\begin{eqnarray}
& &  \partial_{\tau} n_{\bd{k}} ( \tau ; \omega )  
  =  \frac{1}{ V} \sum_{\bd{k}^{\prime}}    
 \int_{ - \infty}^{\infty} \frac{ d \omega^{\prime}}{2 \pi}   
  2 \gamma_{\bd{k} - \bd{k}^{\prime}}^2 D^I_{\bd{k} - \bd{k}^{\prime}} ( \omega - \omega^{\prime} )
 \nonumber
 \\
 & & \times   G^I_{ \bd{k}^{\prime}} ( \tau ;  \omega^{\prime} ) \Bigl\{
 [ 1 + n_{\bd{k}} ( \tau ; \omega ) ]    b ( \omega - \omega^{\prime} )  
 n_{\bd{k}^{\prime}} ( \tau;  \omega^{\prime} ) 
 \nonumber
 \\
 & & \hspace{17mm}
 +   [ 1 +  n_{\bd{k}^{\prime}} ( \tau;  \omega^{\prime} ) ]  
 b (  \omega^{\prime} - \omega )  ] n_{\bd{k}} ( \tau ; \omega ) 
 \Bigr\} .
 \label{eq:kinomega2}
 \end{eqnarray}
Finally, we neglect the damping of intermediate states, which amounts to
approximating
  \begin{equation}
  G^I_{\bd{k}^{\prime}} (\tau ;  \omega^{\prime} )
 \approx 2 \pi \delta ( \omega^{\prime}  - \epsilon_{\bd{k}^{\prime}}   ).
 \end{equation}
Then the $\omega^{\prime}$-integration in Eq.~(\ref{eq:kinomega2}) is trivial
and we obtain for $ n_{\bd{k}} ( \tau ) = n_{\bd{k}} ( \tau ; \epsilon_{\bd{k}} )$
the kinetic equation
 \begin{eqnarray}
  \partial_{\tau} n_{\bd{k}} ( \tau  )  
  & = & \frac{2}{ V} \sum_{\bd{k}^{\prime}}       
  \gamma_{\bd{k} - \bd{k}^{\prime}}^2 D^I_{\bd{k} - \bd{k}^{\prime}} ( 
\epsilon_{\bd{k}} - \epsilon_{\bd{k}^{\prime}} )
 \nonumber
 \\
 &  \times  & \Bigl\{ [ 1 + n_{\bd{k}} ( \tau  ) ] 
 b ( \epsilon_{\bd{k}} - \epsilon_{\bd{k}^{\prime}} )  
 n_{\bd{k}^{\prime}} ( \tau ) 
 \nonumber
 \\
 & &
   -   [ 1 +  n_{\bd{k}^{\prime}} ( \tau) ]   
 [ 1 +   b (      \epsilon_{\bd{k}} -  \epsilon_{\bd{k}^{\prime}}   )  ]
 n_{\bd{k}} ( \tau  )   
 \Bigr\}. \hspace{7mm}
 \label{eq:kinomega3}
 \end{eqnarray}
This is of the form (\ref{eq:master}) with transition rates
 $W_{\bd{k} ,\bd{k}^{\prime}}$ given by Eq.~(\ref{eq:Wk1}).

%

\end{appendix}


\begin{thebibliography}{99}
%
\bibitem{Morawetz04}
K. Morawetz (ed.), {\it{Nonequilibrium Physics  at Short Time Scales}}, (Springer, Berlin, 2004).
%
\bibitem{Polkovnikov11}
A. Polkovnikov, K. Sengupta, A. Silva, and M. Vengalattore,
Rev. Mod. Phys. {\bf{83}}, 863 (2011).
%
\bibitem{Cassing09}
W. Cassing, Eur. Phys. J. Special Topics {\bf{168}}, 3 (2009).
%
\bibitem{Deng10}
H. Deng, H. Haug, and Y. Yamamoto, Rev. Mod. Phys. {\bf{82}}, 1489 (2010).
%
\bibitem{Schoeller09}
H. Schoeller, Eur. Phys. J. Special Topics {\bf{168}}, 179 (2009).
%
\bibitem{Demokritov06}
S.~O. Demokritov, V.~E. Demidov, O.~Dzyapko, G.~A. Melkov, A.~A. Serga,
  B.~Hillebrands, and A.~N. Slavin, Nature {\bf 443}, 430 (2006).
%
%
%
%
%
\bibitem{Dzyapko11}
O. Dzyapko, V. E. Demidov, G. A. Melkov, and S. O. Demokritov,
Phil. Trans. R. Soc. A {\bf{369}}, 3575 (2011).
%
\bibitem{Griffin09}
A. Griffin, T. Nikuni, and E. Zaremba,
{\it{Bose-Condensed Gases at Finite Temperature}},
(Cambridge University Press, Cambridge, 2009).
%
\bibitem{Kamenev11}
  A. Kamenev, {\it{Field Theory of Nonequilibrium Systems}}
(Cambridge University Press, Cambridge, 2011).
%
\bibitem{Berges09}
J. Berges and G. Hoffmeister, Nucl. Phys. B {\bf{813}}, 383 (2009);
J. Berges and D. Sexty,  Phys. Rev. D {\bf{83}}, 085004 (2011);
J. Berges and D. Sexty, Phys. Rev. Lett. {\bf{108}}, 161601 (2012).
%
%
\bibitem{Bagnuls01}
C. Bagnuls and C. Bervillier, Phys. Rep. {\bf{348}}, 91 (2001).
%
\bibitem{Berges02}
J. Berges, N. Tetradis, and C. Wetterich, Phys. Rep. {\bf{363}}, 223 (2002).
%
\bibitem{Kopietz10}
P. Kopietz, L. Bartosch, and F. Sch\"{u}tz, {\it{Introduction
to the Functional Renormalization Group}} (Springer,
Berlin, 2010).
%
\bibitem{Rosten12}
O. J. Rosten,  Phys. Rep. {\bf{511}}, 177 (2012).
%
\bibitem{Metzner12}
W. Metzner, M. Salmhofer, C. Honerkamp, V. Meden, and K. Sch\"{o}nhammer,
Rev. Mod. Phys. {\bf{84}}, 299 (2012).
%
\bibitem{Gasenzer08}
T. Gasenzer and J. M. Pawlowski, Phys. Lett.  B {\bf{670}}, 135 (2008);
T. Gasenzer, S. Ke{\ss}ler, and J. M. Pawlowski, Eur. Phys. J. C {\bf{70}},
 423 (2010).
%
\bibitem{Kloss11}
T. Kloss and P. Kopietz, Phys. Rev. B {\bf{83}}, 205118 (2011).
%
\bibitem{Kennes12}
D. M. Kennes, S. G. Jakobs, C. Karrasch, and V. Meden,
Phys. Rev. B {\bf{85}}, 085113 (2012).
%
\bibitem{Fetter71}
See, for example, A. L. Fetter and J. D. Walecka,
{\it{Quantum Theory of Many-Particle Systems}}
(McGraw-Hill, New York, 1971).
%
\bibitem{Banyai00}
L. B\'{a}nyai, P. Gartner, O. M. Schmitt, and H. Haug,
Phys. Rev. B {\bf{61}}, 8823 (2000).
%
\bibitem{Cherepanov93}
V. Cherepanov, I. Kolokolov, and V. L'vov,
Phys. Rept. {\bf{229}}, 81 (1993).
%
\bibitem{Demidov08a}
V. E. Demidov, O. Dzyapko, 
S. O. Demokritov, G. A. Melkov, and A. N. Slavin,
Phys. Rev. Lett. {\bf{100}}, 047205 (2008).
%
\bibitem{Demidov08b}
V. E. Demidov, O. Dzyapko, M. Buchmeier, T. Stockhoff, G. Schmitz, G. A. Melkov, and
S. O. Demokritov,
Phys. Rev. Lett. {\bf{101}}, 257201 (2008).
%
\bibitem{Kreisel09}
A.~Kreisel, F.~Sauli, L.~Bartosch, and P.~Kopietz,
\newblock Eur. Phys. J. B {\bf 71}, 59 (2009).
%
\bibitem{Kalinikos86}
B. A. Kalinikos and A. N. Slavin, J. Phys. C {\bf{19}}, 7013 (1986); 
B. A. Kalinikos, M. P. Kostylev, N. V. Kozhus, and A. N. Slavin,
J. Phys. Condens. Matter {\bf{2}}, 9861 (1990).
%
\bibitem{Hick10}
J. Hick, F. Sauli, A. Kreisel, and P. Kopietz,
Eur. Phys. J. B {\bf 78}, 429 (2010).
%
\bibitem{Zwanzig01}
Note that our transition rates $W_{\bd{k} , \bd{k}^{\prime}}$
correspond to $W_{\bd{k}^{\prime} , \bd{k} }$ (with interchanged momentum labels)
of Ref.~[\onlinecite{Banyai00}].
We follow here the convention of the textbook by
R. Zwanzig, {\it{Nonequilibrium Statistical Mechanics}}
(Oxford University Press, Oxford, 2001).
%
\bibitem{Negele88}
J. W. Negele and H. Orland, {\it{Quantum Many-Particle Systems}} (Addison-Wesley, 
Redwood City, 1988).
%
\bibitem{Kamenev04}
A. Kamenev, in {\it{Les Houches, Volume Session LX}}, edited by
H. Bouchiat, Y. Gefen, S. Gu\'{e}ron, G. Montambaux, and
J. Dalibard (Elsevier, North-Holland, Amsterdam, 2004).
%
\bibitem{footnoteContour}
As explained, for example by Kamenev \cite{Kamenev11,Kamenev04}, in a
functional integral formalism it is natural
to parametrize the nonequilibrium Green functions in the Keldysh basis and
to work with the three independent Green functions $G^R$, $G^A$ and $G^K$.
The Kadanoff-Baym parametrization corresponds to the contour basis, with
$G^{>} = G^{-+}$ and $G^{<} = G^{+-}$, where
\mbox{$i G^{pp'}_{\bd{k}}(t,t') = \langle a^p_{\bd{k}} ( t )
\bar{a}^{p'}_{\bd{k}} ( t^{\prime} ) \rangle $} with \mbox{$p,p' \in \{
+,-\}$}.
%
\bibitem{Danielewicz84}
P. Danielewicz, Ann. Phys. {\bf{152}}, 239 (1984).
%
\bibitem{Semkat00}
D. Semkat, D. Kremp, and M. Bonitz, J. Math. Phys. {\bf{41}}, 7458 (2000).
%
\bibitem{Garny09}
M. Garny and M. M. M\"{u}ller, Phys. Rev. D {\bf{80}}, 085011 (2009).
%
\bibitem{Rammer07}
J. Rammer, {\it{Quantum Field Theory of Nonequilibrium States}}
(Cambridge University Press, Cambridge, 2007).
%
\bibitem{Wetterich93}
C. Wetterich, Phys. Lett. B {\bf{301}}, 90 (1993).
%
\bibitem{Jacobs10}
 S. G. Jakobs, M. Pletyukhov, and H. Schoeller,
Phys. Rev. B {\bf{81}}, 195109 (2010);
 S. G. Jakobs, M. Pletyukhov, and H. Schoeller,
J. Phys. A: Math. Theor. {\bf{43}},  103001 (2010);
C. Karrasch, M. Pletyukhov, L. Borda, and V. Meden,
Phys. Rev. B {\bf{81}}, 125122 (2010).
%
\bibitem{Schuetz05}
F. Sch\"{u}tz, L. Bartosch, and P. Kopietz, Phys. Rev. B {\bf{72}}, 035107 (2005);
F. Sch\"{u}tz and P. Kopietz, J. Phys. A: Math. Gen. {\bf{39}}, 8205 (2006).
%
\bibitem{Jolicoeur91}
T. Jolicoeur and J. C. Le Guillou, Phys. Rev. B {\bf{44}}, 2403 (1991).
%
\bibitem{Sandweg10}
C. W. Sandweg, M. B. Jungfleisch, V. I. Vasyuchka, A. A. Serga,
P. Clausen, H. Schultheiss, B. Hillebrands, A. Kreisel, and P. Kopietz,
Rev. Sci. Instrum. {\bf{81}}, 073902 (2010).
%
\bibitem{Snoke06}
D. Snoke, Nature {\bf{443}}, 403 (2006).
%
\bibitem{Bugrij07}
A. I. Bugrij and V. M. Loktev, Low Temp. Phys. {\bf{33}}, 37 (2007)
%
\bibitem{Rueckriegel12}
A. R\"{u}ckriegel, A. Kreisel, and P. Kopietz,
 Phys. Rev. B {\bf{85}}, 054422 (2012).
%
\bibitem{Kohn70}
W. Kohn and D. Sherrington, Rev. Mod. Phys. {\bf{42}}, 1 (1970).
%
\bibitem{Jepsen78}
O. Jepsen, J. Madsen, and O. K. Andersen, Phys. Rev. B {\bf{18}}, 605 (1978).
%
\bibitem{Siu01}
G. G. Siu, C. M. Lee, and Y. Liu,
Phys. Rev. B {\bf{64}}, 094421 (2001).
%
\bibitem{Chernyshev12}
A. L. Chernyshev, Phys. Rev. B {\bf{86}}, 060401(R) (2012).
%
\bibitem{Kreisel11}
A. Kreisel, P. Kopietz, P. T. Cong, B. Wolf, and M. Lang,
Phys. Rev. B {\bf{84}}, 024414 (2011).
%
\bibitem{footnoteHaug}
Note that in Ref.~[\onlinecite{Banyai00}]  B\'{a}nyai {\it{et al.}} 
use a decoupling procedure to derive an  equation of motion
for the order parameter $ \psi_0 ( t ) =   \langle a_{\bd{k}=0} ( t ) \rangle$.
It is easy to see that in our notation their order parameter equation
can be written as $i \partial_t \psi_0 ( t ) = \Sigma^R_{ \bd{k} =0 } ( t; \omega =0) 
 \psi_0 ( t )$,  where $\Sigma^R_{ \bd{k}  } (  \tau ; \omega)$ is the Wigner transform
of the retarded self-energy to second order in the
magnon-phonon coupling, see Eq.~(\ref{eq:SigmaRtt}).
This equation can be viewed as a linearized Gross-Pitaevskii equation.
For a consistent description of the condensed phase, the non-linear 
terms in the Gross-Pitaevskii equation should be retained.
These terms arise from magnon-magnon interactions which we have neglected
in our model Hamiltonian defined in Eq.~(\ref{eq:hdef}).
%
\bibitem{Lipavsky86}
P. Lipavsk\'y, V. \u{S}pi\u{c}ka, and B. Velick\'y,
Phys. Rev. B {\bf{34}}, 6933 (1986).
%



%
\end{thebibliography}
\end{document}